\documentstyle[12pt,psfig,axodraw,a4]{article} 
\def\bild#1#2{    
        \vspace*{-5mm}
        \begin{center}
        \begin{math}
        \epsfxsize#2cm
        \epsffile{#1}
        \end{math}
        \end{center}
        }
\newcommand{\vs}{\vspace{-0.25cm}}

\begin{document}

\begin{center}
\Large{\bf Nuclear spin-orbit interaction from \\ chiral pion-nucleon dynamics}

\bigskip

N. Kaiser\\

\bigskip

{\small Physik Department T39, Technische Universit\"{a}t M\"{u}nchen, D-85747
Garching, Germany\\

\smallskip

{\it email: nkaiser@physik.tu-muenchen.de}}

\end{center}

\bigskip

\begin{abstract}
Using the two-loop approximation of chiral perturbation theory, we calculate 
the momentum and density dependent nuclear spin-orbit strength $U_{ls}(p,k_f)$.
This quantity is derived from the spin-dependent part of the interaction energy
$\Sigma_{spin} = {i\over 2}\,\vec \sigma \cdot (\vec q \times\vec p\,) \, 
U_{ls}(p,k_f)$ of a nucleon scattering off weakly inhomogeneous isospin 
symmetric nuclear matter. We find that iterated $1\pi$-exchange generates at 
saturation density, $k_{f0}=272.7\,$MeV, a spin-orbit strength at $p=0$ of 
$U_{ls}(0,k_{f0})\simeq 35$\,MeVfm$^2$ in perfect agreement with the empirical 
value used in the shell model. This novel spin-orbit strength is neither of 
relativistic nor of short range origin. The potential $V_{ls}$ underlying the 
empirical spin-orbit strength $\widetilde U_{ls}= V_{ls} \, r_{ls}^2$ becomes 
a rather weak one, $V_{ls}\simeq 17$\,MeV, after the identification $r_{ls}=
m_\pi^{-1}$ as suggested by the present calculation. We observe however a
strong $p$-dependence of $U_{ls}(p,k_{f0})$ leading even to a sign change above
$p=200\,$MeV. This and other features of the emerging spin-orbit Hamiltonian
which go beyond the usual shell model parametrization leave questions about the
ultimate relevance of the spin-orbit interaction generated by $2\pi$-exchange
for a finite nucleus. We also calculate the complex-valued isovector 
single-particle potential $U_I(p,k_f)+ i\,W_I(p,k_f)$ in isospin asymmetric
nuclear matter proportional to $\tau_3 (N-Z)/(N+Z)$. For the real part we find
reasonable agreement with empirical values and the imaginary part vanishes at
the Fermi-surface $p=k_f$.    
\end{abstract}


\vskip 0.5cm

\section{Introduction and summary}
The introduction of the spin-orbit term into the nuclear single-particle
Hamiltonian by Haxel, Jensen, Suess and Goeppert-Mayer \cite{haxel} in 1949 has
been most decisive for the success of the nuclear shell model. Only with a very
strong and attractive spin-orbit potential one is, for example, able to explain
the observed sequence of so-called magic numbers $\{2,8,20,28,50,82,126,\dots\}
$. The dynamical origin of the strong nuclear spin-orbit force has not been 
fully resolved even up to date. The analogy with the spin-orbit interaction in
atomic physics gave the hint that it could be a relativistic effect. This idea
has lead to the construction of the relativistic (scalar-vector) mean-field
model \cite{walecka}. In this model the spin-independent nuclear potential (of 
approximate depth $-50$\,MeV) results from an almost complete cancelation of a
very strong attraction generated by scalar ($\sigma$-meson) exchange and a 
nearly equally strong repulsion generated by vector ($\omega$-meson) exchange. 
The corresponding spin-orbit term (obtained by a non-relativistic reduction of
the nucleon's Dirac-Hamiltonian) comes out proportional to the coherent sum of
the very large scalar and vector mean-fields. In this sense, the relativistic 
mean-field model gives a simple and natural explanation of the basic features 
of the nuclear shell model potential. Refinements of relativistic mean-field 
models which include additional non-linear couplings of the scalar and vector 
fields or explicitly density-dependent couplings are nowadays widely and 
successfully used for nuclear structure calculations \cite{ring,typel,lenske}.

The nuclear spin-orbit potential arises generally as a many-body effect from 
the underlying spin-orbit term in the (free) nucleon-nucleon scattering 
amplitude. The calculation of the tree level diagrams with one scalar-meson or
one vector-meson exchange between nucleons gives indeed a spin-orbit term in
the NN T-matrix proportional to $1/M^2$, with $M$ denoting the nucleon mass. 
The  nuclear spin-orbit potential corresponding to scalar and vector meson
exchange is therefore obviously a truly relativistic effect. However, the 
quadratic reciprocal scaling of the spin-orbit NN-amplitude with the nucleon 
mass $M$ is not universal, and it changes if one considers the exchange of two
mesons between nucleons, i.e. loop diagrams. For example, irreducible two-pion
exchange gives rise to a spin-orbit term in the NN T-matrix proportional to 
$1/M$ (see eqs.(22,23) in ref.\cite{nnpap1}) and iterated one-pion exchange
produces a spin-orbit term in the NN T-matrix which even scales linearly with
the nucleon mass $M$ (see eq.(33) in ref.\cite{nnpap1}). It is one of the chief
purposes of this paper to investigate in detail the contributions from iterated
one-pion exchange to the nuclear spin-orbit interaction. As already mentioned
the latter arises from the spin-orbit term in the NN T-matrix as a many-body 
effect, e.g. in connection with a nuclear matter calculation. 

In a recent work \cite{nucmat}, we have used chiral perturbation theory for a  
systematic treatment of the nuclear matter many-body problem. In this 
calculation the contributions to the energy per particle, $\bar E(k_f)$,
originate exclusively from one- and two-pion exchange between nucleons and they
are ordered in powers of the Fermi momentum $k_f$ (modulo functions of
$k_f/m_\pi$). It has been demonstrated in ref.\cite{nucmat} that the empirical 
saturation point and the nuclear matter compressibility $K\simeq 255\,$MeV can
be well reproduced at order ${\cal O}(k_f^5)$ in the chiral expansion with 
just one single momentum cut-off scale of $\Lambda \simeq 0.65\,$GeV which 
parametrizes all necessary short range dynamics. Most surprisingly, the
prediction for the asymmetry energy, $A(k_{f0})=33.8\,$MeV, is in very good 
agreement with its empirical value. Furthermore, as a nontrivial fact pure 
neutron matter is predicted to be unbound and the corresponding equation of 
state agrees roughly with that of sophisticated many-body calculations for low 
neutron densities $\rho_n \leq 0.25\,$fm$^{-3}$.  In a subsequent work 
\cite{einpot}, the momentum and density dependent (real) single-particle 
potential $U(p,k_f)$ (i.e. the spin-independent average nuclear mean-field) has
been calculated in the same framework. It was found that chiral $1\pi$- and 
$2\pi$-exchange give rise to a potential depth for a nucleon at the bottom of 
the Fermi sea of $U(0,k_{f0})=-53.2\,$MeV. This value is in very good agreement
with the depth of the empirical optical model potential and the nuclear shell
model potential. Nuclear matter at finite temperatures has been investigated in
the same framework in ref.\cite{liquidgas}. There it was shown that chiral 
$1\pi$- and $2\pi$-exchange reproduce the first-order liquid-gas phase 
transition of isospin symmetric nuclear matter with a realistic value $T_c
\simeq 19\,$MeV of the critical temperature. Our approach to the nuclear
matter problem is in many respects different from most other commonly used 
ones, where one starts from a so-called realistic NN-potential. For example in
the relativistic nuclear matter calculation of ref.\cite{rolf} the $S-$, $P-$ 
and $D$-waves deliver more than 95\% of the potential energy per particle. The
finding that perturbative chiral pion-nucleon dynamics leads already to good 
nuclear matter and single particle properties hints at the fact that the 
detailed NN-interaction is of no more relevance. Fine-tuning of the single 
cut-off scale $\Lambda$ to one nuclear matter observable (the binding energy
per particle $-\bar E(k_{f0})$) is however still necessary in our present 
approach \cite{nucmat}.

It is the purpose of this work to calculate, using the same framework as in
ref.\cite{einpot}, the momentum and density dependent nuclear spin-orbit 
strength $U_{ls}(p,k_f)$. This quantity is derived from the spin-dependent 
part of the interaction energy $\Sigma_{spin} = {i\over 2}\,\vec \sigma \cdot 
(\vec q \times\vec p\,) \, U_{ls}(p,k_f)$ of a nucleon scattering off weakly
inhomogeneous isospin symmetric nuclear matter. We will present here analytical
expressions for the contributions from $1\pi$-exchange and iterated
$1\pi$-exchange to the spin-orbit strength $U_{ls}(p,k_f)$. Furthermore, we 
calculate in isospin asymmetric (homogeneous) nuclear matter the
(complex-valued) isovector single-particle potential $U_I(p,k_f)+i\,W_I(p,k_f)$
accompanied by the isospin double-asymmetry $\tau_3 (N-Z)/(N+Z)$. Our results
can be summarized as follows:
\begin{itemize}
\item[i)] At nuclear matter saturation density, $k_{f0}=272.7\,$MeV,
$1\pi$-exchange and iterated  $1\pi$-exchange generate a spin-orbit strength 
at $p=0$ of $U_{ls}(0,k_{f0})= (0.4+34.7)$\,MeVfm$^2$. This result, which is 
dominated by the contribution of four Hartree-type diagrams, is in perfect 
agreement with the empirical value of the spin-orbit strength $\widetilde 
U_{ls}\simeq 35\,$MeVfm$^2$ \cite{bohr,eder} used in shell-model calculation 
of nuclei. The novel spin-orbit strength found here is neither of relativistic 
nor of short range origin. It is in fact linearly proportional to the nucleon 
mass $M$ and its inherent range is the pion Compton wavelength $m_\pi^{-1}=1.46
\,$fm. The latter feature tempts to an unconventional interpretation of the
strong nuclear spin-orbit interaction. The potential $V_{ls}$ underlying the
empirical spin-orbit strength $\widetilde U_{ls}=V_{ls} \, r_{ls}^2\simeq
35\,$MeVfm$^2$ becomes a rather weak one, namely  $V_{ls} \simeq 17$\,MeV,
after the identification of the effective range $r_{ls}$ with the pion Compton 
wavelength, $r_{ls} =m_\pi^{-1}$, as suggested by the present calculation. 

\item[ii)] We observe however a strong $p$-dependence of $U_{ls}(p,k_{f0})$ 
which leads even to a sign change above $p=200\,$MeV. The calculated spin-orbit
strength $U_{ls}(0,k_{f0})$ depends also strongly on the value of the pion mass
and it shows a pronounced maximum around $m_\pi \simeq 100\,$MeV. A further 
property of the spin-orbit Hamiltonian emerging from our diagrammatic
calculation is that it has (in coordinate space) terms proportional to $\vec
\nabla f(r)$ as well as terms proportional to $f(r)\,\vec \nabla f(r)$ (with
$f(r)=\rho(r)/\rho(0)$ the normalized radial density profile) which get
weighted differently at the surface of a finite nucleus. All such features of
our calculation which go beyond the usual shell model parametrization of the
spin-orbit Hamiltonian leave questions about the ultimate relevance of the
spin-orbit interaction generated by $2\pi$-exchange for a finite nucleus. 
Implementing the present results for $U_{ls}(p,k_f)$ into nuclear structure 
calculations will clarify the role of the nuclear spin-orbit interaction
generated by $2\pi$-exchange.  

\item[iii)] The real part of the isovector single-particle potential $U_I(0,
k_f)$ generated by chiral $1\pi$- and $2\pi$-exchange has a density dependence
very similar to that of the asymmetry energy $A(k_f)$ \cite{nucmat}. At
saturation density, $\rho_0=0.178\,$fm$^{-3}$, we find a repulsive isovector 
potential of $U_I(0,k_{f0})= 47\,$MeV. This prediction is comparable to the
value $U_1 \simeq 33\,$MeV used in shell model calculations \cite{bohr} or the 
value $U_1 \simeq 40\,$MeV deduced from nucleon-nucleus scattering in the
framework of the optical model \cite{hodgson}. The momentum dependence of 
$U_I(p,k_{f0})$ is non-monotonic in the interval $0\leq p\leq k_{f0}$. One 
observes a broad maximum at $p=230\,$MeV where the (real) isovector 
single-particle potential has  increased by about 30\% to the value $63\,$MeV. 
The imaginary part $W_I(p,k_f)$ vanishes (quadratically) at the Fermi-surface
$(p=k_f)$ in accordance with Luttinger's theorem \cite{luttinger}. 
\end{itemize}

\section{General considerations about the spin-orbit term}
Let us begin with recalling the spin-orbit Hamiltonian of the nuclear shell 
model \cite{bohr} which is generally written in the form:
\begin{equation} {\cal H}_{ls}= \widetilde U_{ls} \,{\vec \sigma \cdot\vec \ell
\over 2 r }\, {d f(r) \over d r}\,,\qquad \quad f(r) = {\rho(r)\over \rho(0)}
\,.  \end{equation}
Here, $ \vec \sigma$ is the conventional Pauli spin-vector and $\vec \ell = 
\vec r \times \vec p $ is the orbital angular momentum of a nucleon. $\rho(r)$ 
denotes the radial density distribution of a nucleus (typically parametrized by
a Saxon-Woods function). The empirical value of the nuclear spin-orbit strength
is  $\widetilde U_{ls} \simeq 35\,{\rm MeV fm}^2$ \cite{bohr,eder}. This large
and positive value of $\widetilde U_{ls}$ leads to a strongly attractive
spin-orbit potential acting mainly at the surface of a nucleus.

We wish to calculate the nuclear spin-orbit strength $\widetilde U_{ls}$ (or an
appropriate generalization of it) in the systematic framework of chiral
perturbation theory \cite{nucmat,einpot}. The first observation one makes is
that the spin-orbit interaction vanishes identically in infinite homogeneous 
nuclear matter since there is no preferred center in this system in order to 
define an orbital angular momentum. Therefore one has to generalize the
calculation of the single-particle potential in ref.\cite{einpot} to (at least)
weakly inhomogeneous nuclear matter. The relevant quantity in order to 
extract the nuclear spin-orbit strength is the spin-dependent part of the 
interaction energy of a nucleon scattering off weakly inhomogeneous isospin 
symmetric nuclear matter from initial momentum $\vec p-\vec q/2$ to final
momentum $\vec p+\vec q/2$, which reads: 
\begin{equation}  \Sigma_{spin} = {i\over 2}\, \vec \sigma \cdot (\vec q \times
\vec p\,) \, U_{ls}(p,k_f) \,.\end{equation} 
The (small) momentum transfer $\vec q$ is provided by the Fourier-components 
of the inhomogeneous nuclear matter distribution. The density form factor 
$\Phi(\vec q\,)= \int d^3 r\, e^{-i \vec q \cdot \vec r} f(r)$ plays the
role of a probability distribution of these Fourier-components. The form factor
$\Phi(\vec q\,)$ should be viewed as narrowly peaked function around $\vec 
q=0$ with its Fourier-transform equal to the (slowly varying) density profile, 
$f(r)=(2\pi)^{-3} \int d^3 q\, e^{i \vec q \cdot \vec r} \,\Phi(\vec q\,)$. 
Using this relationship the spin-orbit Hamiltonian ${\cal H}_{ls}$ in eq.(1) 
becomes equal to the Fourier-transform of the product of the density form 
factor and the spin-dependent interaction energy: ${\cal H}_{ls} =(2\pi)^{-3}
\int d^3 q \,e^{i \vec q \cdot \vec r}\,\Phi(\vec q\,)\,\Sigma_{spin}$. 
Consistent with the assumption of a weakly inhomogeneous nuclear matter 
distribution we keep in $\Sigma_{spin}$ only linear terms in $\vec q$ 
corresponding to small spatial density gradients. For practical purposes, this
means that after isolating the proportionality factor $\vec q$ in an explicit 
calculation the momentum and density dependent spin-orbit strength 
$U_{ls}(p,k_f)$ can be finally computed in the limit of homogeneous isospin
symmetric nuclear matter (characterized by its Fermi momentum $k_f$).  
\begin{center}
\SetWidth{2.5}
\begin{picture}(400,120)

\Line(50,0)(50,100)
\Line(43,47)(57,47)
\Line(43,53)(57,53)
\LongArrow(50,25)(50,30)
\LongArrow(50,70)(50,75)
\Text(10,50)[]{$-\Gamma(\vec p_j,\vec q_j)$}
\Text(20,15)[]{$\vec p_j-\vec q_j/2$}
\Text(22,85)[]{$\vec p_j+\vec q_j/2$}
\Vertex(50,0){4}
\Vertex(50,100){4}
\Text(65,0)[]{$\vec r_1$}
\Text(65,100)[]{$\vec r_2$}

\Line(150,0)(150,100)
\Photon(150,50)(200,50){5}{5}
\ArrowArc(230,50)(30,0,180)
\ArrowArc(230,50)(30,180,360)
\Line(253,47)(267,47)
\Line(253,53)(267,53)
\LongArrow(150,25)(150,30)
\LongArrow(150,70)(150,75)
\Vertex(150,50){4}
\Vertex(200,50){4}
\Text(175,65)[]{$\sigma, \omega$}
\Text(125,15)[]{$\vec p-\vec q/2$}
\Text(125,85)[]{$\vec p+\vec q/2$}

\Line(350,0)(350,120)
\DashCArc(350,60)(40,-90,90){6}
\Vertex(350,20){4}
\Vertex(350,100){4}
\Line(343,57)(357,57)
\Line(343,63)(357,63)
\LongArrow(350,39)(350,40)
\LongArrow(350,79)(350,85)

\end{picture}
\end{center}
{\it Fig.1: Left: The double line symbolizes the medium insertion for a weakly
inhomogeneous many-fermion system $\Gamma(\vec p_j,\vec q_j)$ defined by
eqs.(3,4). Middle: The $\sigma$- and $\omega$-exchange Hartree graph. Right: 
The $1\pi$-exchange Fock graph.}

\bigskip

In ref.\cite{einpot} the calculation of the single-particle potential in 
homogeneous isospin symmetric nuclear matter has been organized in the number 
of so-called medium insertions. The latter is a technical notation for the
difference between the vacuum and in-medium nucleon propagator (see eq.(3) in
ref.\cite{nucmat}). In the case of homogeneous nuclear matter a medium 
insertion in a self-energy diagram converts effectively a four-dimensional loop
integration into an integral over a Fermi-sphere of radius $k_f$. The medium
insertion for a (non-relativistic) many-fermion system is generally constructed
from the sum over the occupied  energy eigenstates as \cite{gross}:
\begin{equation} \Gamma(\vec p_j,\vec q_j) = \int d^3r_1 \int d^3r_2\sum_{E\in 
occ} \psi_E(\vec r_2) \psi^*_E(\vec r_1) \, e^{i \vec p_j \cdot (\vec r_1
-\vec r_2)} \,  e^{-{i\over 2} \vec q_j \cdot (\vec r_1+\vec r_2)} \,.
\end{equation} 
The double line in the left picture of Fig.\,1 symbolized this medium insertion
together with the assignment of the out- and in-going nucleon momenta $\vec p_j
\pm\vec q_j/2$. The momentum transfer $\vec q_j$ is provided by the 
Fourier-components of the inhomogeneous matter distribution. Semiclassical
expansions \cite{gross,reif} give for a weakly inhomogeneous and spin-saturated
many-fermion system: 
\begin{equation} \Gamma(\vec p_j,\vec q_j) = \theta( k_f -|\vec p_j|)  \,
\Phi(\vec q_j ) \,\Big\{ 1+ {\cal O}(\vec q_j) \Big\} \,,  \end{equation}  
with $\Phi(\vec q_j) \sim \int d^3 p_j\, \Gamma(\vec p_j,\vec q_j)$ the
density form factor introduced after eq.(2). The subleading ${\cal O}(\vec q_j)
$ term in eq.(4) will in fact never come into play in our diagrammatic
calculation of the spin-dependent interaction energy $\Sigma_{spin}$ to linear
order in $\vec q$.   

As a first check on this formalism we evaluate the $\sigma$- and $\omega
$-exchange Hartree diagram in Fig.\,1. We perform the non-relativistic $1/M
$-expansion of the scalar/vector interaction vertex sandwiched between 
Dirac-spinors for the out- and in-going nucleon (of momentum $\vec p\pm \vec 
q/2\,$) until we obtain the spin-orbit like term $i\, \vec \sigma \cdot (\vec 
q \times \vec p\,)/4M^2$. After that we can take the limit of homogeneous 
nuclear matter and perform the remaining integral over a Fermi-sphere of 
radius $k_f$ which brings one factor of density $\rho$. Putting all pieces 
together we reproduce the familiar result:
\begin{equation} U_{ls}^{(\sigma\omega)}(p,k_f) = {\rho \over 2M^2 } \bigg(
{g^2_{\sigma N} \over m_\sigma^2} +{g^2_{\omega N} \over m_\omega^2} \bigg) \,,
\qquad \rho = {2k_f^3 \over 3\pi^2} \,, \end{equation}
of the relativistic mean-field model \cite{walecka,ring}. The contribution of
the analogous $\sigma$- and $\omega$-exchange Fock diagrams to the nuclear
spin-orbit strength can also be easily calculated with the help of the 
formalism outlined above. We obtain from these Fock diagrams $1/4$ of the 
$\sigma$-exchange contribution and $3/4$ of the $\omega$-exchange contribution 
written in eq.(5). In this work our main interest is focussed on the
nuclear spin-orbit interaction generated by chiral one- and two-pion exchange. 

\section{Diagrammatic calculation of the spin-orbit strength}
In this section we present analytical results for the nuclear spin-orbit
strength $U_{ls}(p,k_f)$ as given by chiral one- and two-pion exchange. We
start with the $1\pi$-exchange Fock graph (last diagram in Fig.\,1). In the
static approximation the product of $\pi N$-interaction vertices $\vec \sigma
\cdot (\vec p - \vec p_1)\,  \vec \sigma \cdot (\vec p - \vec p_1)= (\vec p -
\vec p_1)^2$ is spin-independent. A non-vanishing nuclear spin-orbit strength
comes therefore only as a relativistic $1/M^2$-correction. Isolating the $ i \,
\vec \sigma \times \vec q$ factor from the product of fully relativistic 
pseudovector $\pi N$-interaction vertices, performing the $1/M$-expansion, and
integrating finally over a Fermi-sphere of radius $k_f$, we get from the
$1\pi$-exchange Fock diagram: 
\begin{eqnarray} U^{(1\pi)}_{ls}(p,k_f)&=& {g_A^2m_\pi^3 \over (8\pi f_\pi
M x)^2 }\bigg\{ u^5-ux^4- {4\over 3}u^3x^2+{u\over2}(u^2+5x^2-1) \nonumber \\ 
&& -4x^2\Big[ \arctan(u+x) + \arctan(u-x)\Big] \nonumber \\ && 
+\bigg[(x^2-u^2)^3-{3\over 2}(x^2-u^2)^2 +6x^2 +{1\over 2}\bigg] L(x,u) 
\bigg\} \,. \end{eqnarray} 
Here, we have introduced the dimensionless variables $x=p/m_\pi$ and
$u=k_f/m_\pi$ and the auxiliary function 
\begin{equation}L(x,u)={1\over 4x}\ln{1+(u+x)^2\over1+(u-x)^2}\,.\end{equation}
For the readers convenience we will present the spin-orbit strength at $p=0$ of
each individual diagram in a separate formula, since in most cases the limit 
$x\to 0$ is quite non-trivial. For the $1\pi$-exchange Fock diagram we find 
the simple expression 
\begin{equation} U_{ls}^{(1\pi)}(0,k_f) = {g_A^2 m_\pi^3 \over(4\pi f_\pi M)^2}
\bigg\{{u^3\over 3}+u +{u \over 1+u^2}-2 \arctan u \bigg\} \,,\end{equation}
which gives numerically at saturation density $k_{f0}=272.7\,$MeV only about 
$1.1\%$ of the empirical value $\widetilde U_{ls} \simeq 35\,{\rm MeV fm}^2$. 


\begin{center}
\SetWidth{2.5}
\begin{picture}(400,140)

\Text(20,110)[]{(a)}
\Line(0,0)(0,120)
\DashLine(0,30)(40,30){6}
\DashLine(0,90)(40,90){6}
\Line(40,30)(40,90)
\CArc(40,60)(30,-90,90)
\Line(63,57)(77,57)
\Line(63,63)(77,63)
\Vertex(0,30){4}
\Vertex(0,90){4}
\Vertex(40,30){4}
\Vertex(40,90){4}
\LongArrow(0,58)(0,62)
\LongArrow(40,58)(40,62)

\Text(130,110)[]{(b)}
\Line(110,0)(110,120)
\DashLine(110,30)(150,30){6}
\DashLine(110,90)(150,90){6}
\Line(150,30)(150,90)
\CArc(150,60)(30,-90,90)
\Line(143,57)(157,57)
\Line(143,63)(157,63)
\Vertex(110,30){4}
\Vertex(110,90){4}
\Vertex(150,30){4}
\Vertex(150,90){4}
\Line(103,57)(117,57)
\Line(103,63)(117,63)
\LongArrow(180,62)(180,56)
\LongArrow(110,18)(110,19)
\LongArrow(110,109)(110,110)

\Text(240,110)[]{(c)}
\Line(220,0)(220,120)
\DashLine(220,30)(260,30){6}
\DashLine(220,90)(260,90){6}
\Line(260,30)(260,90)
\CArc(260,60)(30,-90,90)
\Line(283,57)(297,57)
\Line(283,63)(297,63)
\Vertex(220,30){4}
\Vertex(220,90){4}
\Vertex(260,30){4}
\Vertex(260,90){4}
\Line(213,57)(227,57)
\Line(213,63)(227,63)
\LongArrow(260,56)(260,66)
\LongArrow(220,18)(220,19)
\LongArrow(220,109)(220,110)

\Text(350,110)[]{(d)}
\Line(330,0)(330,120)
\DashLine(330,30)(370,30){6}
\DashLine(330,90)(370,90){6}
\Line(370,30)(370,90)
\CArc(370,60)(30,-90,90)
\Line(393,57)(407,57)
\Line(393,63)(407,63)
\Vertex(330,30){4}
\Vertex(330,90){4}
\Vertex(370,30){4}
\Vertex(370,90){4}
\Line(363,57)(377,57)
\Line(363,63)(377,63)
\LongArrow(370,77)(370,80)
\LongArrow(370,38)(370,49)
\LongArrow(330,59)(330,66)
 \end{picture}
\end{center}
{\it Fig.2: The iterated $1\pi$-exchange Hartree graphs. Their isospin factor
in symmetric nuclear matter is 6.}

\bigskip

Next, we come to the evaluation of the four Hartree diagrams of iterated 
one-pion exchange shown in Fig.\,2. We start with the left graph with one
medium insertion, labeled (a). The  relevant $i\,\vec \sigma \times \vec q$ 
prefactor can be isolated already in the first step of the calculation from the
product of $\pi N$-interaction vertices $\vec\sigma\cdot(\vec l-\vec q/2)\,\vec
\sigma \cdot (\vec l + \vec q/2)$ at the open nucleon line. For all remaining
parts of the diagram one can then take the limit of homogeneous nuclear matter
(i.e. $\vec q=0$). Using the analytical results given in section 4.3 of
ref.\cite{nnpap1} for the inner $d^3l$-loop integral we can even perform the
integral over a Fermi-sphere of radius $k_f$. Altogether, we find the following
closed form expression for the spin-orbit strength generated by the Hartree
diagram (a):   
\begin{eqnarray} U^{(a)}_{ls}(p,k_f)&=& {2g_A^4 M m_\pi^2 \over (8\pi x)^3 
f_\pi^4}\bigg\{ 16(u^3+x^3)\arctan(u+x) + 16(x^3-u^3)\arctan(u-x)  \nonumber 
\\ && +u x (7-9u^2-9x^2) + \Big[ 9(u^2-x^2)^2-30u^2-30x^2-7 \Big] x L(x,u)
\bigg\} \,.  \end{eqnarray}
At zero nucleon momentum $(p=0)$ this simplifies to
\begin{equation} U_{ls}^{(a)}(0,k_f) = {g_A^4 M m_\pi^2 \over 32\pi^3f_\pi^4}
\bigg\{4\arctan u-3 u-{u \over 1+u^2} \bigg\} \,.\end{equation}
Note also that $U^{(a)}_{ls}(p,k_f)$ in eq.(9) originates from the real part of
the iterated $1\pi$-exchange spin-orbit NN-amplitude (eq.(33) in
ref.\cite{nnpap1}) evaluated in forward direction and integrated over a
Fermi-sphere of radius $k_f$.  

We continue with the calculation of the Hartree diagrams with two medium
insertions, labeled (b), (c) and (d) in Fig.\,2. In these diagrams, a (small)
momentum transfer $\vec q_{1,2}$ with $\vec q=\vec q_1+\vec q_2$ occurs at each
medium insertion. The important prefactors $i\,\vec \sigma \times \vec q_j$ can
again be isolated in the first step of the calculation from the product of $\pi
N$-interaction vertices at the open nucleon line. After that the vectors $\vec
q_j$ can be set to zero in all remaining components of these diagrams. When 
Fourier-transformed with the density form factors to coordinate space each such
momentum transfer $\vec q_j\,,\,(j =1,2)$ leads to the expression: $(2\pi)^{-6}
\int d^3 q \int d^3 q_j \,e^{i \vec q\cdot \vec r} \, i\vec q_j \,\Phi(\vec
q_j)\,\Phi(\vec q - \vec q_j) = f(r) \, \vec \nabla f(r)$, with $f(r)$ 
the density profile of weakly inhomogeneous nuclear matter. Consistent with the
assumption of a weakly inhomogeneous nuclear matter (i.e. keeping only linear 
terms in small spatial gradients) we can make the replacement: $f(r) \, \vec 
\nabla f(r) \to \vec \nabla f(r)$. We will come back to this point in the next 
section when discussing the results for the spin-orbit strength $U_{ls}(p,k_f)$
as well as their relevance for a finite nucleus. The essential conclusion from
the previous considerations is that for the calculation of $\Sigma_{spin}$ in 
weakly inhomogeneous nuclear matter each momentum transfer $\vec q_j$ can be 
identified with $\vec q$. With the help of this rule and certain techniques to 
reduce six-dimensional principal value integrals over the product of two 
Fermi-spheres of radius $k_f$, we end up with the following result for the 
Hartree diagram (b):
\begin{equation} U^{(b)}_{ls}(p,k_f)= {6g_A^4 M m_\pi^2\over (4\pi f_\pi)^4}
\int_{-1}^1 dy \,{y\over x} \bigg[2u x y +(u^2-x^2y^2)\ln{u+x y\over u-x y} 
\bigg] \bigg[ {3s+2s^3 \over 1+s^2} -3 \arctan s \bigg] \,,
\end{equation} 
with the auxiliary function $s=x y +\sqrt{u^2-x^2+x^2y^2}$. Throughout this
work the momentum $p$ is restricted to the interval $0 \leq p \leq k_f$. At 
zero nucleon momentum $(p=0)$ eq.(11) simplifies to  
\begin{equation} U_{ls}^{(b)}(0,k_f) = {g_A^4 M m_\pi^2 \over (2\pi f_\pi)^4}
\bigg\{2u^2 +{u^2 \over 1+u^2}-3u \arctan u \bigg\} \,.\end{equation}
Similarly, we find for the Hartree diagram  (c):
\begin{equation} U^{(c)}_{ls}(p,k_f)= {12g_A^4 M m_\pi^2\over (4\pi f_\pi)^4}
\int_{-1}^1 dy \, \int_{-x y}^{s-xy} d\xi\,{y(xy+\xi)^4\over x[1+(xy+\xi)^2]^2}
\,\bigg[2u \xi +(u^2-\xi^2) \ln{u+\xi\over u-\xi} \bigg] \,,\end{equation} 
\begin{eqnarray} U^{(c)}_{ls}(0,k_f)&=& {g_A^4 M m_\pi^2\over (2\pi f_\pi)^4}
\int_0^u d\xi\, {\xi^4 \over (1+\xi^2)^3} \bigg\{ -u(7+3\xi^2) 
\nonumber \\ && +(5u+7\xi+u \xi^2+3\xi^3) \bigg[ 1+{u-\xi\over 2u} \ln{u+\xi 
\over u-\xi} \bigg] \bigg\} \,, \end{eqnarray}
and for the Hartree diagram (d):
\begin{equation} U^{(d)}_{ls}(p,k_f)= {3g_A^4 M m_\pi^2\over (2\pi f_\pi)^4}
\int_0^u d\xi\,{\xi^2 \over x^3} \int_{-1}^1 dy \,\bigg[{\xi y\over 2}
\ln{|x+\xi y|\over |x-\xi y|} -x \bigg] \bigg[ {3\sigma+2\sigma^3 \over 
1+\sigma^2} -3 \arctan \sigma \bigg] \,, \end{equation} 
with the auxiliary function $\sigma =\xi y +\sqrt{u^2-\xi^2+\xi^2y^2}$.
The limit $x\to 0$ of eq.(15) is quite non-trivial since it requires a careful
treatment of the singular function Re$(y+i\,0)^{-2}$. The following
representation is manifestly free of singularities,  
\begin{eqnarray} U^{(d)}_{ls}(0,k_f)&=& {g_A^4 M m_\pi^2\over (2\pi f_\pi)^4}
\int_0^u d\xi \int_{-1}^1 {dy\over y^2} \,\bigg\{ {3\sigma+2\sigma^3 \over 
1+\sigma^2} -3 \arctan \sigma -  {2\xi y(u^2-\xi^2)^2 \over (1+ u^2-
\xi^2)^2} \nonumber \\ && +(1+y^2) \bigg[ 3 \arctan\sqrt{u^2-\xi^2}-{3+2u^2
-2 \xi^2 \over 1+u^2-\xi^2}\sqrt{u^2-\xi^2} \bigg] \bigg\} \,. \end{eqnarray}
Note also the similarity of the expressions eqs.(11,13,15) with the three-body 
single-particle potential given in eq.(11) of ref.\cite{einpot}.

\bigskip

\begin{center}
\SetWidth{2.5}
\begin{picture}(400,140)

\Text(20,135)[]{(e)}
\Line(0,0)(0,140)
\DashCArc(0,52.5)(35,-90,90){6}
\DashCArc(0,87.5)(35,-90,90){6}
\Vertex(0,17.5){4}
\Vertex(0,52.5){4}
\Vertex(0,87.5){4}
\Vertex(0,122.5){4}
\Line(-7,67)(7,67)
\Line(-7,73)(7,73)
\LongArrow(0,106)(0,111)
\LongArrow(0,38)(0,41)

\Text(130,135)[]{(f)}
\Line(110,0)(110,140)
\DashCArc(110,52.5)(35,-90,90){6}
\DashCArc(110,87.5)(35,-90,90){6}
\Vertex(110,17.5){4}
\Vertex(110,52.5){4}
\Vertex(110,87.5){4}
\Vertex(110,122.5){4}
\Line(103,102)(117,102)
\Line(103,108)(117,108)
\Line(103,32)(117,32)
\Line(103,38)(117,38)
\LongArrow(110,69)(110,76)

\Text(240,135)[]{(g)}
\Line(220,0)(220,140)
\DashCArc(220,52.5)(35,-90,90){6}
\DashCArc(220,87.5)(35,-90,90){6}
\Vertex(220,17.5){4}
\Vertex(220,52.5){4}
\Vertex(220,87.5){4}
\Vertex(220,122.5){4}
\Line(213,102)(227,102)
\Line(213,108)(227,108)
\Line(213,67)(227,67)
\Line(213,73)(227,73)
\LongArrow(220,39)(220,41)

\Text(350,135)[]{(g)}
\Line(330,0)(330,140)
\DashCArc(330,52.5)(35,-90,90){6}
\DashCArc(330,87.5)(35,-90,90){6}
\Vertex(330,17.5){4}
\Vertex(330,52.5){4}
\Vertex(330,87.5){4}
\Vertex(330,122.5){4}
\Line(323,67)(337,67)
\Line(323,73)(337,73)
\Line(323,32)(337,32)
\Line(323,38)(337,38)
\LongArrow(330,107)(330,111)

\end{picture}
\end{center}
{\it Fig.3: The iterated $1\pi$-exchange Fock graphs. Their isospin factor
in symmetric nuclear matter is $-3$.}

\bigskip

Next, we come to the evaluation of the four iterated $1\pi$-exchange Fock 
diagrams shown in Fig.\,3. We start with diagram (e) with one medium insertion.
Let us first consider the $\pi N$-interaction vertices. The spin-dependent 
part of the matrix-product $\vec \sigma \cdot \vec a \,\, \vec \sigma \cdot\vec
b \,\,\vec\sigma\cdot \vec a \,\,\vec\sigma \cdot \vec b$ is $2i\, \vec a \cdot
\vec b \,\, \vec \sigma\cdot (\vec a \times\vec b\,)$. In the case of graph (e)
one makes the assignment: $\vec a = \vec l+\vec Q $ and $\vec b = \vec l$, with
$\vec l$ the loop momentum and $\vec Q = \vec p_1-\vec p$, where $\vec p$ 
belongs to the external nucleon line and $\vec p_1$ to the internal nucleon 
line carrying the medium insertion. The important proportionality factor $\vec
q$ is now not produced by the $\pi N$-interaction vertices as it was the case
for the Hartree diagrams in Fig.\,2. In order to isolate the factor $\vec q$ we
combine the previously mentioned $\vec a \cdot \vec b$ term with the nucleon 
energy denominator (which in the actual calculation results from the $dl_0
$-loop integration) and employ the identity: $\vec l \cdot (\vec l+ \vec Q)\,[
\vec l \cdot (\vec l+\vec Q+\vec q\,)]^{-1}=1-\vec l \cdot\vec q \,\,[\vec l
\cdot (\vec l+\vec Q+\vec q\,)]^{-1}$. The term coming along with $1$ in this 
decomposition finally loop-integrates to zero. From the second term one can
now easily isolate the factor $\vec q$ and take the limit to homogeneous
nuclear matter of all remaining factors of the diagram. Putting all pieces
together we end up with the following representation for the nuclear spin-orbit
strength generated by the iterated $1\pi$-exchange Fock diagram (e): 
\begin{eqnarray} U^{(e)}_{ls}(p,k_f)&=& {3g_A^4 M m_\pi^2 \over (8\pi x)^3 
f_\pi^4}\bigg\{  u x (u^2+x^2)-{1\over 2}(u^2-x^2)^2 \ln{u+x\over u-x}
 \nonumber \\ && + \int_{(u-x)/2}^{(u+x)/2}d\xi \,{ [u^2-(2\xi+x)^2]
[u^2-(2\xi-x)^2] \over 4\xi^2 (1+2\xi^2)}  \nonumber \\ && \times\Big[(1+4\xi^2
) \arctan 2\xi -4\xi^2(1+\xi^2) \arctan \xi \Big] \bigg\} \,, \end{eqnarray}
\begin{equation} U_{ls}^{(e)}(0,k_f) = {g_A^4 M m_\pi^2 \over (4\pi)^3f_\pi^4}
\bigg\{u +{1\over 2+u^2} \bigg[ {u^2 \over 2}(4+u^2)\arctan {u\over 2}-2(1+u^2)
 \arctan u\bigg] \bigg\} \,.\end{equation}
Note also that $U^{(e)}_{ls}(p,k_f)$ in eq.(17) originates from the real part
of the iterated $1\pi$-exchange spin-orbit NN-amplitude (eq.(33) in
ref.\cite{nnpap1}) evaluated in backward direction and integrated over a
Fermi-sphere of radius $k_f$. 

We continue with the computation of the Fock diagrams with two medium 
insertion, labeled (f) and (g) in Fig.\,3. Diagram (f) with a symmetrical 
arrangement of the two medium insertions leads to the following contribution to
the nuclear spin-orbit strength: 
\begin{eqnarray} U^{(f)}_{ls}(p,k_f)&=& {3g_A^4 M m_\pi^2\over (4\pi f_\pi)^4}
\bigg\{ {1\over 4x^4} \bigg[\Big(x^2(4u^2-3x^2)-(1+u^2)^2\Big) L(x,u) 
\nonumber \\ && + u(1+u^2+3x^2)-2x^2\Big(\arctan(u+x)+\arctan(u-x)\Big)\bigg] 
\nonumber \\ &&  \times \bigg[ u(1+u^2+x^2)-[1+(u+x)^2][1+(u-x)^2] L(x,u)\bigg]
\nonumber \\ && +\int_{-1}^1 dy\, \int_{-1}^1 dz \,{y|z| \,\theta(y^2+z^2-1)
\over x|y| \sqrt{y^2+z^2-1}} \Big[ \ln(1+s^2)-s^2 \Big] \Big[ t-\arctan t
\Big]\bigg\} \,, \end{eqnarray} 
with the auxiliary function $t=x z +\sqrt{u^2-x^2+x^2z^2}$. At zero nucleon
momentum $(p=0)$ the lengthy expression eq.(19) simplifies drastically to
\begin{equation} U^{(f)}_{ls}(0,k_f)= {g_A^4 M m_\pi^2\over (2\pi f_\pi)^4}
\, {u^6\over 3(1+u^2)^2} \,. \end{equation} 
Note the leading $\rho^2$-behavior of the expressions in eqs.(12,20) derived 
from diagrams with two medium insertions. The results eqs.(8,10,18) belonging 
to diagrams with one medium insertion, on the other hand, show a leading linear
dependence on the density $\rho \sim u^3$. 

Finally, we have to evaluate the last two topologically distinct Fock diagrams
in Fig.\,3. Since they contribute equally to the nuclear spin-orbit strength 
we have given both diagrams the same label (g). In order to avoid very lengthy 
formulas we split their contribution to $U_{ls}(p,k_f)$ into a "factorizable" 
$(g')$ and a "non-factorizable" part $(g'')$. Technically these two pieces are
distinguished by the feature whether the nucleon propagator in the denominator
can be canceled by terms from the product of $\pi N$-interaction vertices
in the numerator, or not. We find from the iterated $1\pi$-exchange Fock
diagrams (g) with two medium insertions the following "factorizable"
contribution to the nuclear spin-orbit strength:  
\begin{eqnarray} U^{(g')}_{ls}(p,k_f)&=& {3g_A^4 M m_\pi^2\over (4\pi f_\pi)^4
x^3} \int_0^u d\xi \, \bigg\{ {3ux\over 2\xi^2}(1+u^2)(1+x^2) -{ux\over2} 
(12\xi^2 +1+x^2)\nonumber \\ && +4\xi^2\Big[\arctan(u+\xi)+\arctan(u-\xi) \Big]
\Big[x+(x^2-1-\xi^2)L(\xi,x) \Big]  \nonumber \\ &&  +\bigg[ {3\xi^4\over 2} 
(4u^2+5x^2-3) +{\xi^2 \over 2}(5+10u^2-4x^2-14u^2x^2-x^4)-6\xi^6-u^2x^4
\nonumber \\ && +{1\over 2}(5+3u^4+3x^2-u^4x^2)+2u^2+x^4-3u^2x^2+{3\over2\xi^2}
(1+u^2)^2(1+x^2)^2 \bigg] \nonumber \\ && \times L(\xi,x) L(\xi,u) +  \bigg[   
6 \xi^4 +{ \xi^2 \over 2} (x^2-12u^2-3)+ (1+x^2)(u^2-1)\nonumber \\ &&
-{3 \over 2 \xi^2} (1+x^2)(1+u^2)^2 \bigg] x L(\xi,u) + \bigg[ 6\xi^4 +{ \xi^2 
\over 2}( 13-15x^2)\nonumber \\ && +{1 \over 2}(x^4+u^2x^2 +3x^2-3u^2-2) 
 -{3 \over 2 \xi^2}(1+x^2)^2(1+u^2) \bigg]u L(\xi,x) \bigg\}\,, \end{eqnarray} 
which turns at zero nucleon momentum $(p=0)$ into the form: 
\begin{eqnarray} U^{(g')}_{ls}(0,k_f)&=& {2g_A^4 M m_\pi^2\over(4\pi f_\pi)^4} 
\int_0^u d\xi \, {\xi^2 \over (1+\xi^2)^2} \bigg\{ u(3u^2-9\xi^2 -17) 
 + 4(3+\xi^2)  \Big[ \arctan(u+\xi) \nonumber \\ && +\arctan(u-\xi) 
\Big] + (9\xi^4 -6u^2 \xi^2 +18\xi^2 -3u^4 -26u^2 -7) L(\xi,u) \bigg\} \,. 
\end{eqnarray}
The analytical evaluation of the "non-factorizable" parts from diagrams (g) 
terminates with two non-elementary integrations and we find the following 
representation for their contribution to the nuclear spin-orbit strength: 
\begin{eqnarray} U^{(g'')}_{ls}(p,k_f)&=& {3g_A^4 M m_\pi^2\over (4\pi f_\pi)^4
} \int_0^u d\xi \, {\xi^2 \over x^3} \int_{-1}^1 dy \bigg\{ 8\Big[\sigma - 
\arctan\sigma \Big] \Big[x-L(\xi,x)\Big]  \nonumber \\ &&  +\bigg[ {1\over
2} \ln(1+ \sigma^2) +4\xi y \arctan\sigma +2u^2-2\xi^2 -{5\over 2 } \sigma^2
\bigg] \ln{|x+\xi y|\over |x-\xi y|}\nonumber \\ &&  + \bigg[ 4\xi y \arctan
\sigma+ {1\over 2} (1+\xi^2-x^2) \ln(1+ \sigma^2)  +2u^2-2\xi^2  \nonumber \\
&& +{\sigma^2 \over 2 }(x^2-5-\xi^2) \bigg]  {1\over R } \ln{ |x R+
(x^2-1-\xi^2)\xi y| \over |x R+ (1+\xi^2-x^2)\xi y|} \bigg\}\,, \end{eqnarray}
with the auxiliary function $R=\sqrt{(1+x^2-\xi^2)^2+4\xi^2(1-y^2)}$. At zero
nucleon momentum $(p=0)$ eq.(23) turns into the (singularity free) form:   
\begin{eqnarray} U^{(g'')}_{ls}(0,k_f)&=& {4g_A^4 M m_\pi^2\over (4\pi f_\pi)^4
} \int_0^u d\xi \, {\xi \over (1+\xi^2)^2} \int_{-1}^1 {dy \over y^2} \bigg\{ 
2\xi(1+\xi^2-2y^2) \Big[\arctan\sigma - \sigma \Big] \nonumber \\ &&+y\bigg[ 
\ln(1+\sigma^2) -\ln(1+ u^2-\xi^2)-2\xi y \sigma + {2\xi^2 (1+\xi^2)(u^2-\xi^2)
\over 1+u^2-\xi^2 }\bigg] \nonumber \\ && +2\xi(1+\xi^2)(1+y^2) \Big[ \sqrt{u^2
-\xi^2} -\arctan\sqrt{u^2-\xi^2} \Big]\bigg\}\,.  \end{eqnarray} 
Note again the similarity of the expressions in eqs.(19,21,23) with the 
three-body potential given in eq.(13) of ref.\cite{einpot}. We also like to
emphasize that the techniques used to reduce six-dimensional principal value
integrals over the product of two Fermi-spheres of radius $k_f$ to at most
double integrals have been checked rigorously in ref.\cite{einpot} via the 
Hugenholtz-van-Hove theorem.  

Let us end this section with power counting considerations for the
spin-dependent interaction energy  $\Sigma_{spin}$. Counting the quantities 
$\vec p,\, \vec q,\, k_f$ and $m_\pi$ collectively as small momenta, we deduce 
that the relativistic $1/M^2$-correction from the $1\pi$-exchange Fock graph 
is of fifth order, while all contributions from iterated $1\pi$-exchange are of
fourth order in small momenta. Irreducible $2\pi$-exchange gives also rise to 
a fifth order contribution to the (real part of the) single-particle potential
(see eq.(16) in ref.\cite{einpot}). To the spin-orbit NN-amplitude irreducible
$2\pi$-exchange contributes as a (higher-order) relativistic $1/M$-correction
(see eqs.(22,23) in ref.\cite{nnpap1}) and it will therefore enter $\Sigma_
{spin}$ at sixth order in small momenta. We have also checked that the
relativistic $1/M$-corrections to iterated $1\pi$-exchange start to contribute
to $\Sigma_{spin}$ first at sixth order in small momenta. From all that we can
conclude that the present calculation of the nuclear spin-orbit strength
$U_{ls}(p,k_f)$ is complete up-to-and-including third order in small momenta. 
Strictly speaking, this counting argument applies only to the long-range 
effects induced by chiral $1\pi$- and $2\pi$-exchange since it does not cover 
the possible short-range contribution, $U^{(short)}_{ls}(p,k_f) =C_{ls}\,\rho$,
which is of course also of third order in small momenta. From this point of
view the $\sigma\omega $-exchange term eq.(5) (together with the Fock
contributions) provides a model for the a priori undetermined low-energy
constant $C_{ls}$. In order to learn about the convergence of the chiral
expansion of the nuclear spin-orbit strength $U_{ls}(p,k_f)$ one should
calculate the contributions from irreducible $2\pi$-exchange of fourth order in
small momenta. Work along this line is in progress. Furthermore, we note that 
spin-orbit strength generated by the three-body force diagrams in Fig.\,2 of 
ref.\cite{pieper} is (formally) of higher order in small momenta. These should 
also be evaluated with the formalism introduced in section 2.

\begin{table}
\begin{center}
\begin{tabular}{|c|ccccccccc|}
    \hline

    Diagram & $1\pi$-Fock & (a) & (b) & (c) & (d) & (e)  & (f) & (g') & (g'') 
     \\  \hline     $U_{ls}(0,k_{f0})$ 
& 0.40 &$-$52.96 & 37.35 & 41.11 & 12.57 & 30.33 & 14.69 & $-$44.50 & $-$3.85 
\\    \hline
$U_{ls}(k_{f0},k_{f0})$
& $-$0.28 &$-$40.05 & 31.63 & 13.95 & $-$35.35 & 25.63 & 4.35 & $-$29.05 & 8.04
 \\    \hline
  \end{tabular}
\end{center}
\end{table}
\vspace{-0.7cm}
{\it Table\,1: Contributions of individual diagrams to the nuclear spin-orbit
strength $U_{ls}(p,k_{f0})$ at $p=0$ and at $p=k_{f0}=272.7\,$MeV. The units
are MeVfm$^2$.}

\bigskip

\subsection{Results}
For the numerical evaluation of the nuclear spin-orbit strength $U_{ls}(p,k_f)$
we use consistently the same parameters as in our previous works 
\cite{nucmat,einpot}. We choose the value $g_A=1.3$ for the nucleon axial 
vector coupling constant. The weak pion decay constant has the value $f_\pi = 
92.4\,$MeV and $M=939\,$MeV and $m_\pi= 135\,$MeV are the masses of the nucleon
and the (neutral) pion, respectively.

In the second row of Table\,1, we present numerical values for the
contributions of individual diagrams to the spin-orbit strength 
$U_{ls}(0,k_{f0})$ at nuclear matter saturation density $k_{f0}=272.7\,$MeV. As
expected the relativistic $1/M^2$-correction from the $1\pi$-exchange Fock
graph is a very small $1.1\%$ effect. The contributions of individual iterated
$1\pi$-exchange diagrams are surprisingly large. In several cases they even
exceed the empirical value $\widetilde U_{ls} \simeq 35\,$MeVfm$^2$ in
magnitude and moreover they are of alternating signs. The basic reason for
these large values is the large scale enhancement factor $M$ (the nucleon mass)
entering the iterated $1\pi$-exchange. The proportionality factor $M$ stems 
from the energy denominator of such second-order diagrams which is a difference
of small nucleon kinetic energies. Adding up the entries in the second row of 
Table\,1 one gets $U_{ls}(0,k_{f0}) =35.1\,{\rm MeVfm}^2$, which is in perfect 
agreement with the empirical value of the nuclear spin-orbit strength 
$\widetilde U_{ls}\simeq 35\,$MeVfm$^2 $ \cite{bohr,eder}. The predicted total 
sum is dominated by the contribution $U^{(H)}_{ls}(0,k_{f0}) =38.1\,{\rm 
MeVfm}^2$ of the four Hartree diagrams (a), (b), (c) and (d) (see Fig.\,2). 
Interestingly, the same feature, namely the numerical suppression of the 
iterated $1\pi$-exchange Fock diagrams against the Hartree diagrams, holds also
for the (spin-independent) nuclear mean-field $U(p,k_f)$ studied in 
ref.\cite{einpot}. The novel spin-orbit strength generated (almost completely)
by iterated  $1\pi$-exchange is neither of relativistic nor of short range
origin. It is in fact linearly proportional to the nucleon mass $M$ and its
inherent range is the pion Compton wavelength $m_\pi^{-1}=1.46\,$fm. The latter
feature tempts to an unconventional interpretation of the strong nuclear
spin-orbit interaction. The potential $V_{ls}$ underlying the empirical
spin-orbit strength $\widetilde U_{ls}= V_{ls} \, r_{ls}^2\simeq 35\,$MeVfm$^2$
becomes a rather weak one, namely $V_{ls} \simeq 17$\,MeV, after the
identification of the effective range $r_{ls}$ with the pion Compton
wavelength, $r_{ls} =m_\pi^{-1}=1.46\,$fm, as suggested by the present 
calculation.   

As a side remark we consider $U_{ls}(0,k_f)$ in the chiral limit $m_\pi=0$. In 
that case all occurring integrals can be performed analytically and we find 
the following simple expression: 
\begin{equation} U_{ls}(0,k_f)_{|m_\pi=0}= {g_A^4 M k_f^2\over(4\pi f_\pi)^4}
\bigg( {4\over 3} -{\pi^2 \over 2} \bigg) + {g_A^2 k_f^3\over 3(4\pi f_\pi 
M)^2}\,, \end{equation}
which gives at $k_{f0}=272.7\,$MeV the negative value $-15.0\,$MeVfm$^2$. 

In Fig.\,4, we show the pion mass dependence of the nuclear spin-orbit strength
$U_{ls}(0,k_{f0})$ at nuclear matter saturation density (and zero nucleon
momentum). One observes a sign change and a pronounced maximum at $m_\pi \simeq
100\,$MeV. 
\bigskip
 
\bild{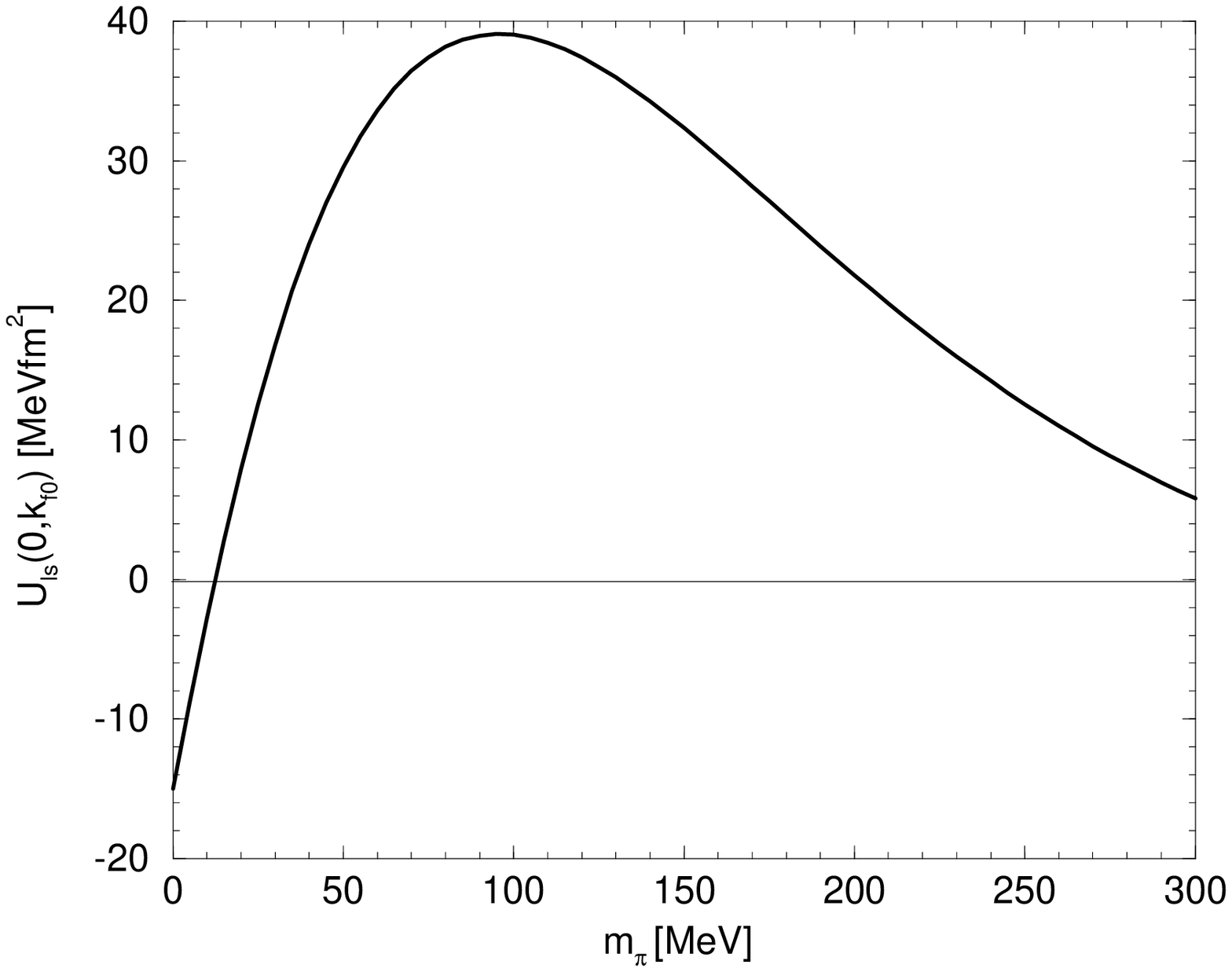}{14}
\vspace{-0.8cm}
{\it Fig.\,4: The pion mass dependence of the nuclear spin-orbit strength
$U_{ls}(0,k_{f0})$ at saturation density $k_{f0}=272.7$\,MeV.}

\bigskip 
   
\bild{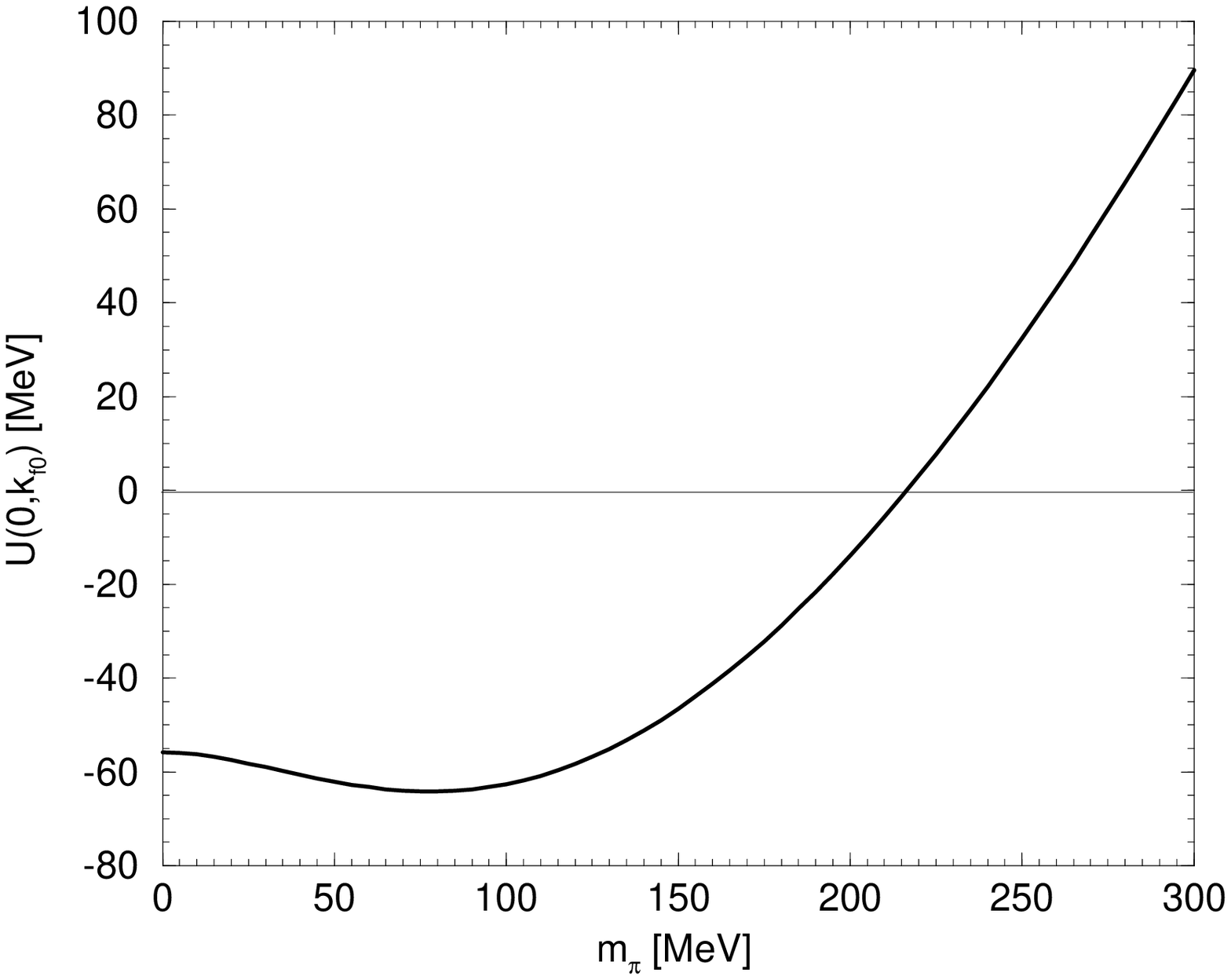}{14}
\vspace{-0.8cm}
{\it Fig.\,5: The pion mass dependence of the depth of the spin-independent
single-particle potential $U(0,k_{f0})$ \cite{einpot} at saturation density 
$k_{f0}=272.7$\,MeV.}

\bigskip This strong $m_\pi$-dependence of $U_{ls}(0,k_{f0})$ has its origin
in the alternating signs of the large contributions from individual diagrams as
well as their own specific $m_\pi$-dependence.  For comparison, we show in
Fig.\,5 the pion mass dependence of the depth of the (real) single-particle 
potential $U(0,k_{f0})$ calculated in ref.\cite{einpot}. One observes a weak
variation of the potential depth $U(0,k_{f0})$ by at most $10\%$ if $m_\pi$
runs from zero (chiral limit) to  the physical value $m_\pi= 135\,$MeV. With
increasing pion mass the attractive nuclear mean-field generated by chiral
$2\pi$-exchange gets however soon lost and it turns into repulsion above
$m_\pi\simeq215\,$MeV. 
 
In Fig.\,6, we show by the full line the dependence of the calculated
nuclear spin-orbit strength $U_{ls}(0,k_f)$ on the nucleon density $\rho=2
k_f^3/3\pi^2$. One observes in the region $\rho \leq 0.4\,$fm$^{-3}$ an 
approximate linear growth of $U_{ls}(0,k_f)$ with the density as it is, for
example, known from $\sigma$- and $\omega$-exchange (see eq.(5)). 

In a finite nucleus the spin-orbit force acts mainly on the surface where the
density gradients are largest and the density has dropped to about half of the
central density. The replacement $f(r) \,\vec \nabla f(r)\to {1\over 2}\vec
\nabla f(r)$ (instead of $f(r) \,\vec \nabla f(r)\to \vec \nabla f(r)$ valid 
for weakly inhomogeneous nuclear matter) describes then more realistically the
situation for a finite nucleus. The dashed line in Fig.\,6 shows the spin-orbit
strength which results if the contributions from the diagrams with two medium 
insertion (b), (c), (d), (f) and (g) are weighted with a factor $1/2$. This 
different weighting leads to a substantial reduction of the total spin-orbit 
strength such that only about $18\%$ of the empirical value $\widetilde U_{ls}
\simeq 35\,$MeVfm$^2$ are left at nuclear matter saturation density $\rho_0 =
0.178\,$fm$^{-3}$.

\bigskip

\bild{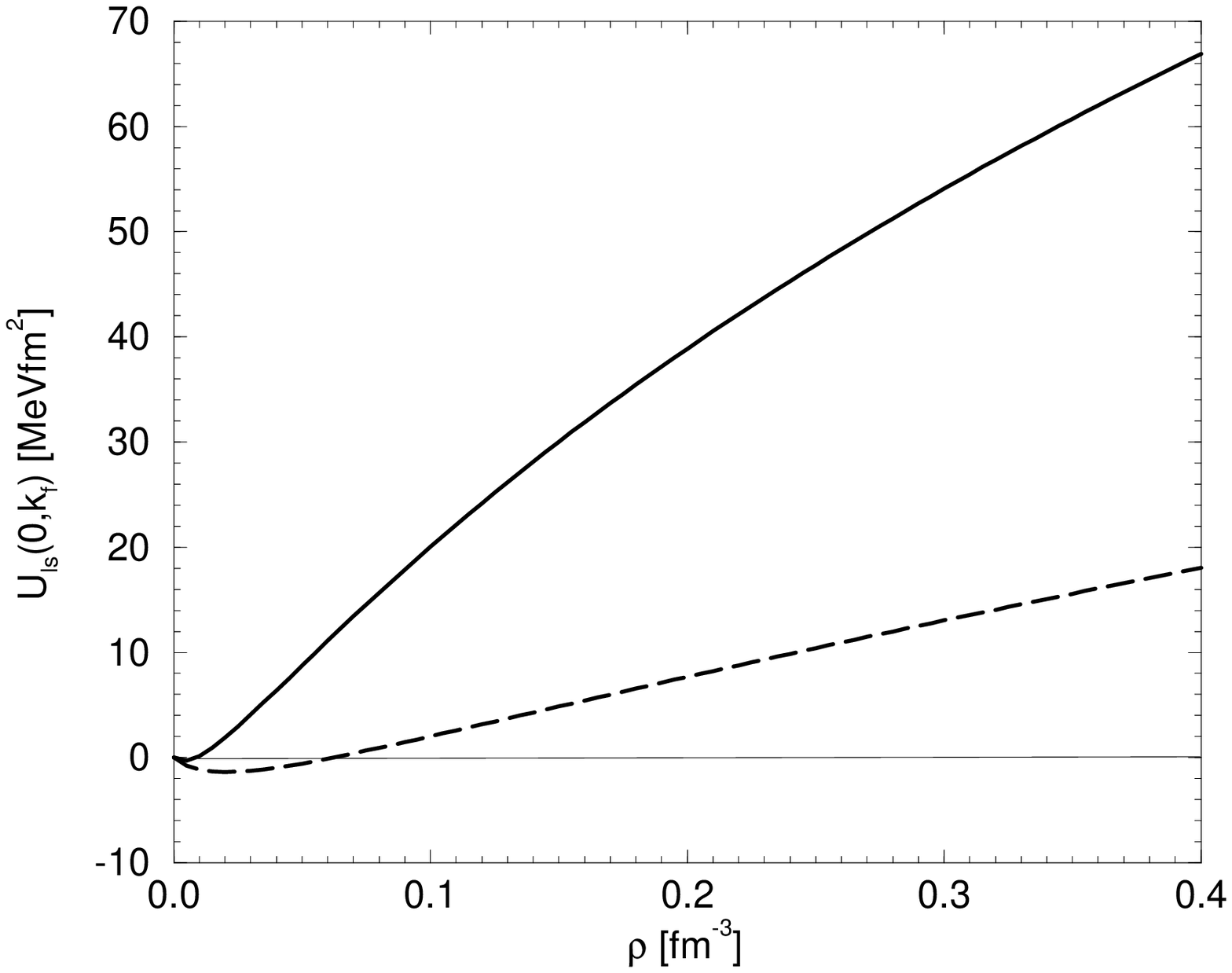}{14}
\vspace{-0.8cm}
{\it Fig.\,6: The full line shows the nuclear spin-orbit strength $U_{ls}(0,k_f
)$ at zero nucleon momentum $(p=0)$ versus the nucleon density 
$\rho=2k_f^3/3\pi^2$. If the diagrams with two medium insertions are weighted
with a factor $1/2$ the dashed line results.}  

\bigskip

In Fig.\,7, we show by the full line the dependence of the calculated nuclear 
spin-orbit strength $U_{ls}(p,k_{f0})$ at saturation density $k_{f0}=272.7 
$\,MeV on the nucleon momentum $p$ for $0\leq p\leq k_{f0}$. One observes a 
very strong $p$-dependence which leads even a sign change of $U_{ls}(p,k_{f0})
$ above $p=200\,$MeV. Again, this strong $p$-dependence of $U_{ls}(p,k_{f0})$
has its origin in the alternating signs of the large contributions from
individual diagrams as well as their own specific $p$-dependence. The numerical
values in the second and third row of Table\,1 indicate how these contributions
from individual diagrams change with the nucleon momentum from $p=0$ to $p=
k_{f0}$. One also should keep in mind that the scale relevant for momentum 
dependences is here set by the pion mass, $m_\pi=135\,$MeV. Note that the ratio
$p/m_\pi$ changes by two units from $p=0$ to $p=k_{f0}$. Furthermore, the 
dashed line in Fig.\,7 corresponds to the weighting of diagrams with two medium
insertions with a factor 1/2. It shows the same strong $p$-dependence including
a sign change as the full line in Fig.\,7.   

\bigskip

\bild{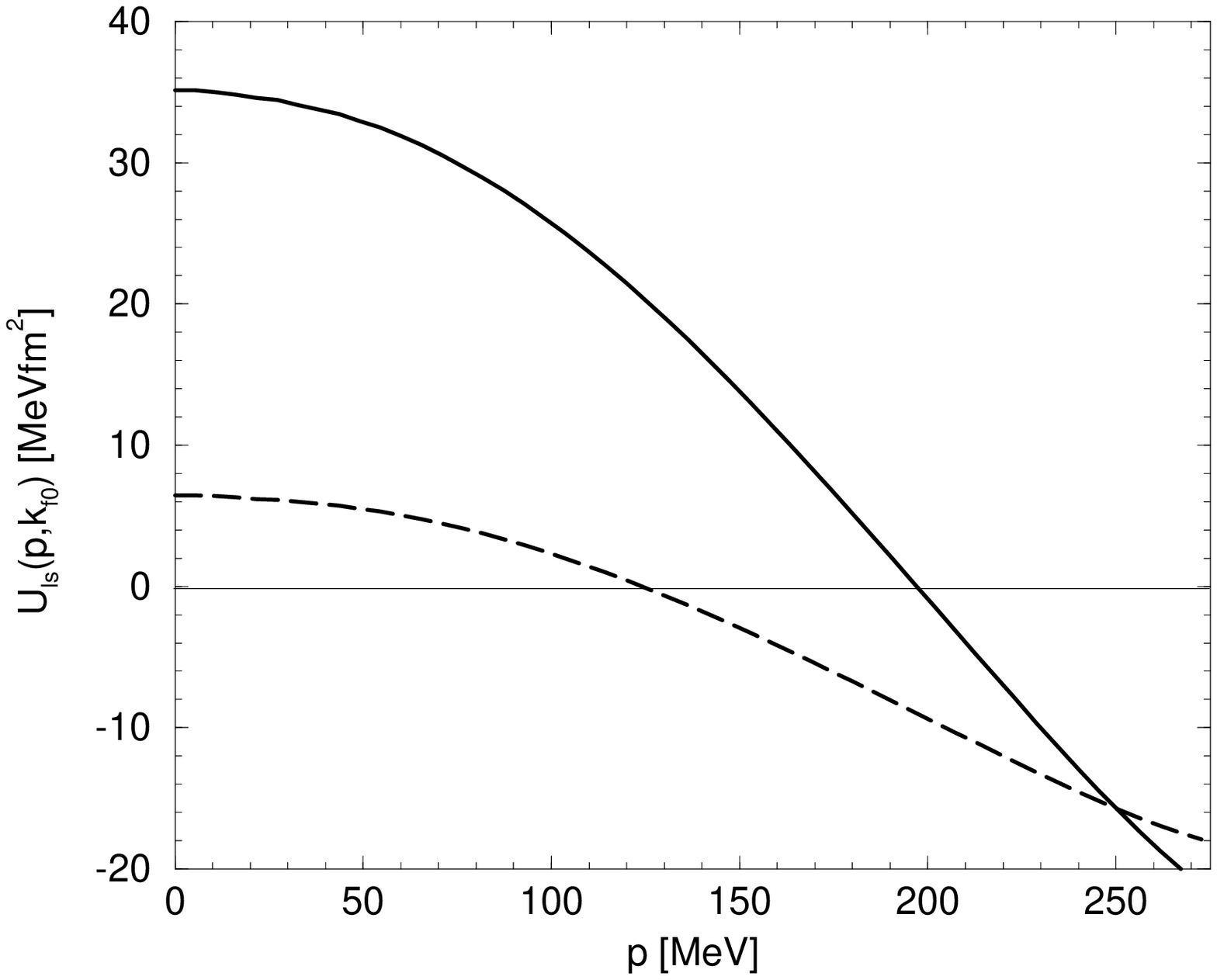}{14}
\vspace{-0.8cm}
{\it Fig.\,7: The momentum dependence of the nuclear spin-orbit strength
$U_{ls}(p,k_{f0})$ at saturation density $k_{f0}=272.7$\,MeV. The full (dashed)
line corresponds to the weighting of diagrams with two medium insertions with a
factor 1 (1/2).}

\bigskip

Let us briefly summarize our results. We have demonstrated here that the
nuclear spin-orbit interaction is not necessarily a relativistic effect. 
Iterated $1\pi$-exchange (i.e. loosely speaking, the $1\pi$-exchange spin-spin
and tensor force in second order) generates large nuclear spin-orbit terms
which in fact scale linearly with the nucleon mass $M$. For weakly
inhomogeneous nuclear matter and small nucleon momenta ($p\leq 70\,$MeV) the
spin-orbit strength from iterated $1\pi$-exchange agrees (at saturation
density) well with the empirical shell model value $\widetilde U_{ls}\simeq
35\,$MeVfm$^2$. The strong $p$-dependence of $U_{ls}(p,k_{f0})$ (including a
sign change) and the different weighting of $\vec \nabla f(r)$ and $f(r) \,
\vec \nabla f(r)$ at the nuclear surface leave however questions about the 
ultimate relevance of this $2\pi$-exchange spin-orbit interaction for real
nuclear structure. Nuclear structure calculation which use the calculated
spin-orbit strength $U_{ls}(p,k_f)$ as input are necessary in order to clarify
the role of the spin-orbit interaction generated by $2\pi$-exchange.           

\section{Isovector single-particle potential}
In this section we generalize our previous calculation \cite{einpot} of the
single-particle potential to isospin asymmetric (homogeneous) nuclear
matter. Any relative excess of neutrons over protons in the nuclear medium
leads to a different single-particle potential for a proton and a neutron. This
fact is expressed by the following decomposition of the single-particle 
potential in isospin asymmetric nuclear matter:     
\begin{equation} U(p,k_f)+i\, W(p,k_f) - \Big[U_I(p,k_f) +i\, W_I(p,k_f)\Big]
\, \tau_3 \, \delta +{\cal O}(\delta^2)  \,. \end{equation}
Here, $U(p,k_f)+i\, W(p,k_f)$ is the (complex) single-particle potential in 
isospin symmetric nuclear matter. The term linear in the isospin asymmetry 
parameter $\delta=(N-Z)/(N+Z)$ defines the (complex) isovector single-particle 
potential $U_I(p,k_f)+i\, W_I(p,k_f)$, and $\tau_3 \to \pm 1$ for a proton or 
a neutron. The situation of an isospin asymmetric nuclear medium is realized 
by the simple substitution:     
\begin{equation} \theta(k_f-|\vec p_j|) \quad \to \quad {1+\tau_3 \over 2}\,  
\theta(k_p-|\vec p_j|) +{1-\tau_3 \over 2}\, \theta(k_n-|\vec p_j|)\,,
\end{equation}
in the medium insertion eq.(4). Here, $k_{p,n} = k_f(1\mp \delta)^{1/3}$ denote
the (different) Fermi momenta of protons and neutrons. Differences in
comparison to the calculation of $U(p,k_f)+i\, W(p,k_f)$ in ref.\cite{einpot} 
occur only with respect to isospin factors and the radii of the Fermi-spheres,
$k_{p,n} =  k_f(1\mp \delta)^{1/3}$. In practise the isovector single-particle
potential $U_I(p,k_f)+i\, W_I(p,k_f)$ is obtained by differentiating the
$\tau_3$-components of the diagrammatic expressions with respect to $\delta$ 
at $\delta=0$.

\subsection{Real part} 
Without going into further technical details we enumerate now the individual 
contributions to the real part of the isovector single-particle potential 
$U_I(p,k_f)$.

\medskip

\noindent i) $1\pi$-exchange Fock diagram in Fig.\,1 including the relativistic
$1/M^2$-correction:
\begin{eqnarray}U^{(1\pi)}_I(p,k_f) &=& {g_A^2m_\pi^3u^2\over3(4\pi f_\pi)^2 } 
\bigg\{ 2L(x,u)-2u +{m_\pi^2 \over M^2} \bigg[ u(u^2+x^2) \nonumber \\ &&
-(u^2+x^2)L(x,u)
+{u(u^2-x^2)^2 \over 2[1+(u+x)^2][1+(u-x)^2]} \bigg] \bigg\} \,, \end{eqnarray}
with $x=p/m_\pi$ and $u=k_f/m_\pi$. The auxiliary function $L(x,u)$ has been 
defined in eq.(7).   

\medskip

\noindent ii) Iterated $1\pi$-exchange Hartree graphs in Fig.\,2:
\begin{equation}U^{(a)}_I(p,k_f) = {g_A^4 M m_\pi^4 u^2 \over 48\pi^3 f_\pi^4} 
\bigg\{\Big({u\over x}-1\Big)  \arctan(u-x)-\Big({u\over x}+1\Big)\arctan(u+x)
+{5\over 2} L(x,u) \bigg\}  \,, \end{equation}

\begin{eqnarray}U^{(b)}_I(p,k_f) &=& {2g_A^4 M m_\pi^4 \over 3(4\pi f_\pi)^4} 
\int_{-1}^1 d y \, \bigg\{  2u^2 \ln{u+x y\over u-x y }\bigg[ {2s^2+s^4 \over
1+s^2} -2\ln(1+s^2)\bigg] \nonumber \\ && - {s^5 s' \over (1+s^2)^2} \bigg[
2uxy +(u^2-x^2y^2) \ln{u+x y\over u-x y }\bigg] \bigg\}  \,, \end{eqnarray}
with $s'= u\,\partial s/\partial u$ and the auxiliary function $s$ has been 
defined after eq.(11). 
\begin{eqnarray}U^{(c)}_I(p,k_f) &=& {4g_A^4 M m_\pi^4 \over 3(4\pi f_\pi)^4 } 
\int_{-1}^1 dy \int_{-xy}^{s-xy} d\xi\, {(xy+\xi)^5 \over s[1+(xy+\xi)^2]^2} 
\bigg\{ s'(2\xi +x y) \bigg[\xi \ln{u+\xi \over u-\xi}-2u\bigg] \nonumber \\ 
&& -{2 s' \over 1+(xy+\xi)^2}\bigg[ 2u \xi +(u^2-\xi^2) \ln{u+\xi \over u-\xi}
\bigg]- (2s+s'\,) u^2 \ln{u+\xi \over  u-\xi} \bigg\} \,, \end{eqnarray}
\begin{eqnarray}U^{(d)}_I(p,k_f) &=& {4g_A^4 M m_\pi^4  \over 3(4\pi f_\pi)^4} 
\int_0^u d\xi \, {\xi^2 \over x}-\!\!\!\!\!\!\int_{-1}^1 dy \,\bigg\{  
{2 x \xi y \over x^2-\xi^2 y^2} \bigg[ {2\sigma^2+\sigma^4\over 1+\sigma^2 }
-2\ln(1+\sigma^2)\bigg] \nonumber \\ && + \bigg[{\sigma^2\over (1+\sigma^2)^2}
(5\sigma^4 +9\sigma^2 +6 -4\sigma^3 \sigma')-6 \ln(1+\sigma^2)\bigg]\ln{|x+ 
\xi y|\over |x- \xi y|} \bigg\}\,, \end{eqnarray}
with $\sigma'= u\,\partial \sigma /\partial u$ and the auxiliary function
$\sigma$ has been defined  after eq.(15). The symbol $-\!\!\!\!\!\int_{-1}^1 
dy$ stands for a principal value integral.  

\medskip

\noindent iii) Iterated $1\pi$-exchange Fock graphs in Fig.\,3:
\begin{eqnarray}U^{(e)}_I(p,k_f) &=& {5g_A^4 M m_\pi^4 u^2 \over 6(4\pi)^3
f_\pi^4 } \bigg\{-2u + \int_{(u-x)/2}^{(u+x)/2} {d\xi \over x(1+2\xi^2)}
\nonumber \\ && \times \Big[ (1+4\xi^2)\arctan 2\xi- (1+8\xi^2+8\xi^4) \arctan
\xi \Big]  \bigg\} \,,  \end{eqnarray} 
\begin{eqnarray}U^{(f)}_I(p,k_f) &=& {g_A^4 M m_\pi^4 \over 3(4\pi f_\pi)^4 } 
\bigg\{- {u \, G(x,u) \over 4 x^2} \,{\partial G(x,u)\over \partial u}
\nonumber \\ && + \int_{-1}^1 dy \int_{-1}^1 dz \,{ y z\,\theta(y^2+z^2-1) 
\over |y z| \sqrt{y^2+z^2-1}} \,{ s^3 s'\over 1+s^2} \Big[ t^2- \ln(1+t^2) 
\Big]  \bigg\} \,, \end{eqnarray}
with the auxiliary function $G(x,u)= u(1+u^2+x^2) - [1+(u+x)^2][1+(u-x)^2]
\,L(x,u)$. 
\begin{eqnarray}U^{(g')}_I(p,k_f) &=& {g_A^4 M m_\pi^4 \over3(4\pi f_\pi)^4 }
\bigg\{ G(u,u) \Big[(1+x^2-u^2) L(x,u)-u\Big] \nonumber \\ && + 5u \int_0^u 
d\xi \,\bigg[ {1\over \xi}(1+x^2-\xi^2) L(x,\xi) -1 \bigg] \,{\partial G(\xi,u)
\over \partial u}  \bigg\} \,, \end{eqnarray}
\begin{eqnarray}U^{(g'')}_I(p,k_f) &=& {g_A^4 M m_\pi^4\over(4\pi f_\pi)^4} 
\int_0^u d\xi \, {\xi^2 \over 3x} -\!\!\!\!\!\!\int_{-1}^1 dy \bigg\{
\bigg[ {\sigma^2 \over 1+\sigma^2}( 3+5\sigma^2+8\sigma \sigma' ) - 3\ln(1+
\sigma^2) \bigg] \nonumber \\ && \times \bigg[ \ln{|x+ \xi y|\over |x- \xi y|} 
+ {1 \over R} \ln{|x R +(x^2-\xi^2-1) \xi y|\over|x R +(\xi^2+1-x^2) \xi y|}
\bigg] +\xi \Big[\ln(1+\sigma^2)-  \sigma^2\Big] \nonumber \\ && \times 
\bigg[{R' \over R^2}  \ln{|x R +(x^2-\xi^2-1) \xi y|\over|x R +(\xi^2+1-x^2) 
\xi y|} + { y(1-x^2+3\xi^2) -x R' \over R[ x R+(x^2-\xi^2-1) \xi y]}\nonumber 
\\ && +{ y(1-x^2+3\xi^2) +x R' \over R[ x R +(\xi^2+1-x^2) \xi y]} - {2 x y 
\over x^2 - \xi^2 y^2}  \bigg] \bigg\}\,, \end{eqnarray}
with $R'= \partial R /\partial \xi$ and the auxiliary function $R$ has been 
defined after eq.(23). 

\medskip

\noindent iv) Irreducible $2\pi$-exchange Hartree and Fock graphs: 
\begin{equation}U^{(2\pi)}_I(p,k_f) = {m_\pi^5 u^2 \over 18x (4\pi f_\pi)^4 }
\bigg\{ J\Big({u+x\over 2}\Big)-J\Big({u-x\over 2}\Big)\bigg\}\,,\end{equation}
\begin{eqnarray} J(\xi) &=& 3(1+2g_A^2+5g_A^4) \ln^2(\xi+\sqrt{1+\xi^2})
\nonumber \\ && + 2\Big[ 5+26g_A^2-79 g_A^4 +2\xi^2 ( 1+10 g_A^2-59 g_A^4)
\Big] \xi \sqrt{1+\xi^2}\ln(\xi+\sqrt{1+\xi^2})\nonumber \\ && +(17+242
g_A^2-787 g_A^4) \xi^2 -(3+14 g_A^2+15g_A^4) \xi^4 \nonumber \\ && +\Big[ 60(1
+6g_A^2-15 g_A^4) \xi^2 +4(1+10 g_A^2-59g_A^4) \xi^4 \Big] \ln{m_\pi \over
2\Lambda} \,. \end{eqnarray} 

\medskip

\noindent v) Power divergences specific for cut-off regularization: 
\begin{equation}U^{(\Lambda)}_I(p,k_f) = { 2\Lambda\, k_f^3 \over 3 (4\pi
f_\pi)^4 } \bigg[ 26 g_A^4 M +5(3g_A^2+1)(1-g_A^2) \Lambda \bigg] \,.
\end{equation} 
The term linear linear in the cut-off $\Lambda$ stems from iterated
$1\pi$-exchange with a contribution of the Hartree diagram (a) and the Fock
diagram (e) in the ratio $8:5$. The term quadratic in the cut-off $\Lambda$, on
the other hand, originates from irreducible $2\pi$-exchange. Note that the
$p$-independent contribution to $U_I(p,k_f)$ in eq.(39) is just twice its
contribution to the asymmetry energy $A(k_f)$ (see eq.(29) in
ref.\cite{nucmat}). This relative factor of 2 is typical for a momentum 
independent NN-contact interaction, to which the power divergences are
completely equivalent, as emphasized in ref.\cite{nucmat}. We use consistently
the value $\Lambda =646.3\,$MeV $\simeq 7f_\pi$ of the cut-off scale which has
been  fine-tuned in ref.\cite{nucmat} to the binding energy per particle,
$-\bar E(k_{f0})=15.26\,$MeV.   

\medskip

\bild{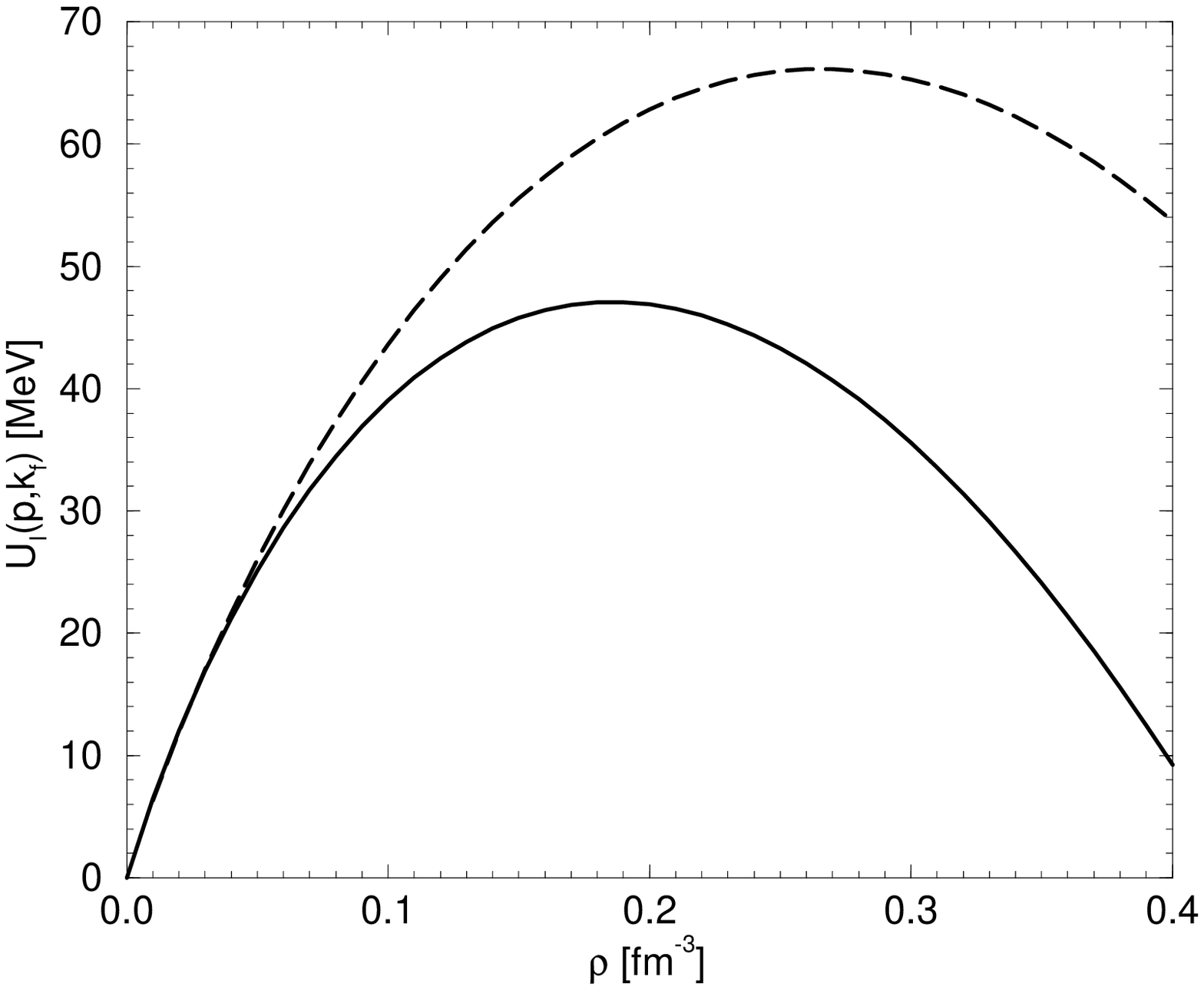}{14}
\vspace{-0.8cm}
{\it Fig.\,8: The real part of the isovector single-particle potential 
$U_I(p,k_f)$ in isospin asymmetric nuclear matter versus the nucleon density
$\rho =2k_f^3/3\pi^2$. The full (dashed) line corresponds to $p=0$ ($p=k_f$).}

\bigskip

In Fig.\,8, we show by the full line the total (real) isovector single particle
potential $U_I(0,k_f)$ of our calculation for a nucleon at rest ($p=0$) as a
function of the nucleon density $\rho= 2k_f^3/3\pi^2$. The shape of this curve
is very similar to the asymmetry energy $A(k_f)$ (see Fig.\,7 in ref.\cite{
nucmat}). In comparison to $A(k_f)$ the scale on the ordinate is stretched by 
a factor of about 1.4.  Interestingly, the (real part of the) isovector 
single-particle potential $U_I(0,k_f)$ has its maximum close to the saturation 
density $\rho_0=0.178\,$fm$^{-3}$. The actual value at that point is 
$U_I(0,k_f)=47.0$\,MeV. This prediction is comparable to the value $U_1 \simeq 
33\,$MeV \cite{bohr} used in shell model calculations or the value $U_1 \simeq 
40\,$MeV  \cite{hodgson} deduced from nucleon-nucleus scattering in the 
framework of the optical model. The dashed line in Fig.\,8 shows the density 
dependence of the (real) isovector single-particle potential $U_I(k_f,k_f)$ at
the Fermi surface $p=k_f$. At that point the (real) isovector single-particle 
potential comes out always more repulsive than at $p=0$. Note also that the
(possibly unrealistic) downward bending branches of the curves in Fig.\,8 
start at densities higher than those relevant for conventional nuclear physics.

Furthermore, we show in Fig.\,9 the momentum dependence of the (real) isovector
single-particle potential $U_I(p,k_{f0})$ at saturation density $k_{f0}=272.7
$\,MeV. The $p$-dependence of $U_I(p,k_{f0})$ is non-monotonic in the interval 
$0\leq p\leq k_{f0}$. One observes a broad maximum at $p=230\,$MeV where 
$U_I(p,k_{f0})$ has increased by about 30\% to the value $62.9\,$MeV.
Note that in comparison to the spin-orbit strength $U_{ls}(p,k_{f0})$ shown in
Fig.\,7 the $p$-dependence of the real part of the isovector single-particle 
potential  $U_I(p,k_{f0})$ is very moderate. Most useful would be nuclear
structure calculations using the calculated isovector single particle
potential $U_I(p,k_f)$ as input.  

\medskip

\bild{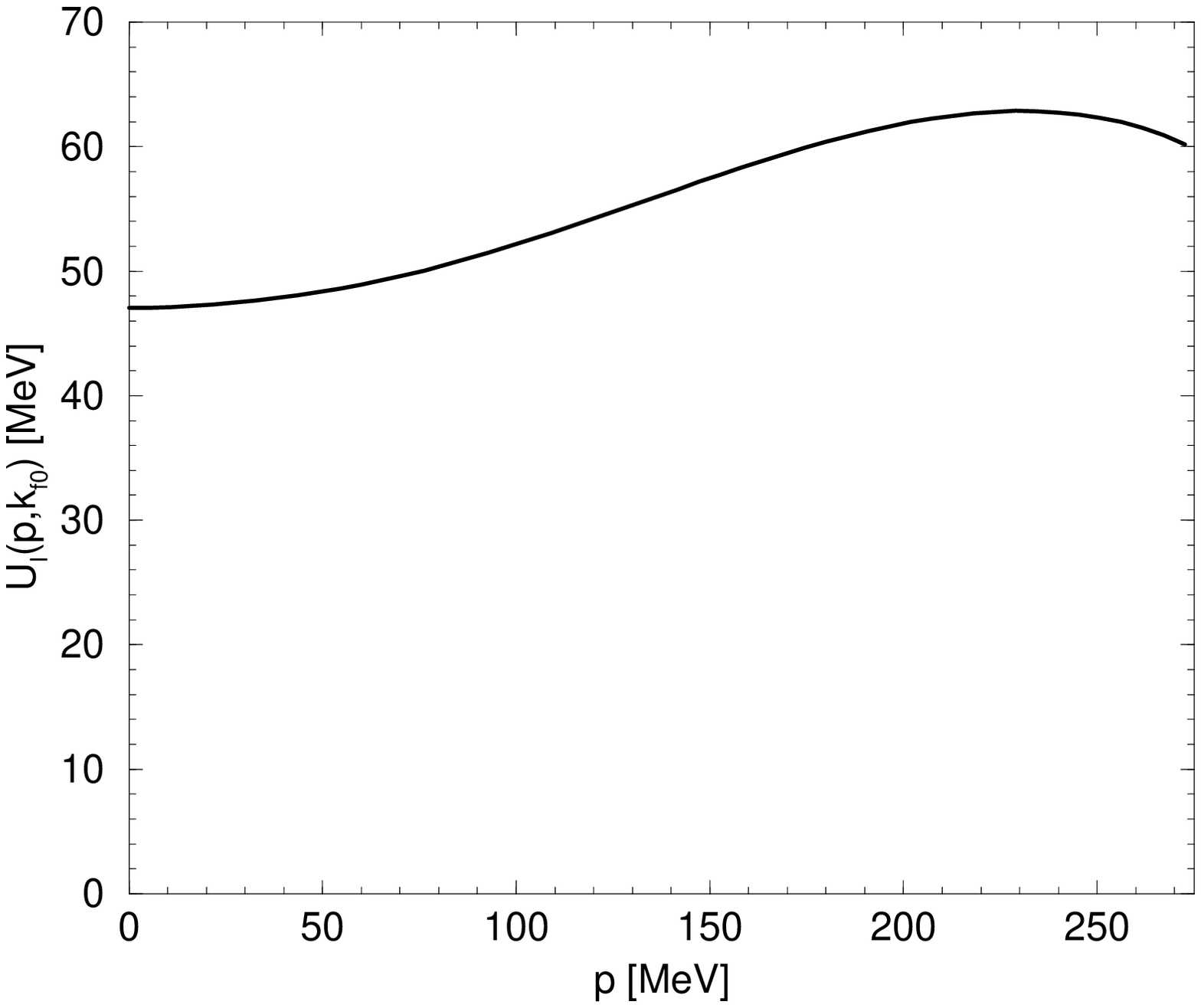}{14}
\vspace{-0.8cm}
{\it Fig.\,9: The momentum dependence of the real part of the isovector 
single-particle potential $U_I(p,k_{f0})$ at saturation density
$k_{f0}=272.7$\,MeV.} 

\bigskip

\subsection{Imaginary part}
In this subsection, we discuss the imaginary part $W_I(p,k_f)$ of the isovector
single-particle potential. According to eq.(26), it splits and shifts the 
half-width of a neutron-hole state and a proton-hole state in the Fermi-sea
(with momentum $0\leq p \leq k_f$) by the amount $\pm\delta \,W_I(p,k_f)$. 
Within the present calculation the imaginary part $W_I(p,k_f)$ arises entirely
from iterated one-pion exchange. It is advantageous to present analytical
formulas only for complete classes of diagrams. We find from the iterated
$1\pi$-exchange Hartree diagrams in Fig.\,2 (including the analogous graph with
three medium insertions) the following contribution to the imaginary isovector 
single-particle potential: 
\begin{eqnarray}W^{(H)}_I(p,k_f) &=& {g_A^4 M m_\pi^4 \over 192\pi^3 f_\pi^4} 
\Bigg\{2u^4-10u^2-{14\over 3}u^2x^2 + 8u^2 \ln(1+4x^2)\nonumber \\ && +{5u^2
\over x} \Big[ \arctan(u-x)- \arctan(u+x)+2 \arctan2x \Big] \nonumber \\ && -
2u^2 \Big(1+{u\over x}\Big) \ln[1+(u+x)^2]+2u^2 \Big({u\over x}-1\Big) 
\ln[1+(u-x)^2] \nonumber \\ && +\int_{-1}^1 dy \,\bigg\{ 2u^2 \bigg[ {2s^2+s^4
\over 1+s^2}-2\ln(1+s^2) \bigg] +{(x^2-u^2) s^5 s'\over 2(1+s^2)^2} \nonumber
\\ && +\int_0^u d\xi\,\xi^2 \bigg[\delta(x-\xi|y|)\bigg(2\ln(1+\sigma^2)-
{2\sigma^2 +\sigma^4 \over 1+\sigma^2}\bigg) + \theta(x-\xi|y|) \nonumber \\ &&\times{1\over x} \bigg({\sigma^2  \over 1+\sigma^2}(6+9\sigma^2+5\sigma^4
-4 \sigma^3 \sigma'\,)-6\ln(1+\sigma^2)\bigg)\bigg] \bigg\}\Bigg\}  \,.
\end{eqnarray}
The iterated $1\pi$-exchange Fock diagrams in Fig.\,3 (again including the 
analogous graph with three medium insertions) lead, on the other hand, to the
following expression:
\begin{eqnarray}W^{(F)}_I(p,k_f) &=& {\pi g_A^4 M m_\pi^4\over 3(4\pi f_\pi)^4}
\Bigg\{u^2 \ln(1+4x^2)+{u^2\over x} \arctan2x- u^2(2+5u^2+3x^2) \nonumber \\ &&
+ \int_0^1 dz \, {u^2[4x^2z^2-\ln(1+ 4x^2z^2)] \over \sqrt{(1+x^2-u^2)^2+4(u^2-
x^2z^2)}}+{5u^2 \over x}\int_{(u-x)/2}^{(u+x)/2} d\xi \, {1+4\xi^2 \over 
1+2\xi^2}\nonumber \\ && \times  \ln(1+4\xi^2)+ \int_{-1}^1 dy \, 
\Bigg\{\int_{-1}^1 dz \, {\theta(1-y^2-z^2)\over \pi \sqrt{1-y^2-z^2}}\,{s^3 s'
\over 1+s^2} \Big[ t^2-\ln(1+t^2)\Big]\nonumber \\ && +\int_0^u d\xi\, {\xi^2 
\over x} \,\Bigg[ \theta(x-\xi |y|) {\xi R'\over R^2}\Big(\sigma^2-\ln(1+
\sigma^2)\Big) +\bigg(1-{1\over R}\bigg) \bigg[\theta(x-\xi |y|)\nonumber \\ &&
\times  \bigg({\sigma^2\over 1+\sigma^2} \Big[\sigma \sigma' \,[8-10\,\theta(
\xi-x)] +3+5 \sigma^2 \Big] -3 \ln(1+\sigma^2) \bigg) \nonumber \\ && + x\, 
\delta(x-\xi |y|) \Big(\ln(1+\sigma^2)-\sigma^2\Big) -{ 2u^2 \sigma_x^3
\, \theta(x-\xi) \over (\sigma_x- \xi y) (1+\sigma_x^2) } \bigg]\Bigg]
\Bigg\}\Bigg\}  \,, \end{eqnarray} 
with the auxiliary function $\sigma_x = \xi y +\sqrt{u^2-x^2+\xi^2 y^2}$. If
the delta-function $\delta(x-\xi|y|)$ is used to eliminate the $dy$-integration
in eqs.(40,41) the remaining $d\xi$-integral extends over the restricted 
region $x\leq \xi \leq u$. The sum of both contibutions eqs.(40,41) evaluated 
at zero nucleon momentum $(p=0)$ can even be written as a closed form
expression: 
\begin{equation} W_I(0,k_f) = {g_A^4 M m_\pi^4 u^2\over 384\pi^3 f_\pi^4}
\bigg\{ {9u^6+40u^4+27u^2\over 2(1+u^2)^2}- {12u^4+40u^2+27\over
(1+u^2)(2+u^2)} \ln(1+u^2) \bigg\} \,. \end{equation} 

Finally, we show in Fig.\,10 the momentum dependence of the imaginary isovector
single-particle potential $W_I(p,k_{f0})$ at saturation density $k_{f0}=272.7
$\,MeV. The associated value at zero nucleon momentum, $W_I(0,k_{f0})=26.7
\,$MeV, agrees within $10\%$ with the isoscalar half-width $W(0,k_{f0})=29.7
\,$MeV found in ref.\cite{einpot}. As a consequence of the decreasing phase 
space available for redistribution of a nucleon-hole state's energy, the curve 
in Fig.\,10 drops with momentum $p$ and $W_I(p,k_{f0})$ reaches zero at the 
Fermi-surface $p=k_{f0}$. The exact vanishing of $W_I(p,k_f)$ at the 
Fermi-surface $p=k_f$ is even separately true for the class of iterated
$1\pi$-exchange Hartree diagrams and the class of iterated $1\pi$-exchange Fock
diagrams. The conditions $W^{(H)}_I(k_f,k_f)=0$ and $W^{(F)}_I(k_f,k_f)=0$
serve as an excellent (analytical and numerical) check on the involved
calculations leading to eqs.(40,41). 

In summary, we find that the predictions from chiral $1\pi$- and 
$2\pi$-exchange for the real part of the isovector single-particle potential 
$U_I(p,k_f)$ agree fairly well with empirical values. The calculated imaginary
part $W_I(p,k_f)$ fulfills the constraints imposed by Luttinger's theorem
\cite{luttinger}.

\medskip

\bild{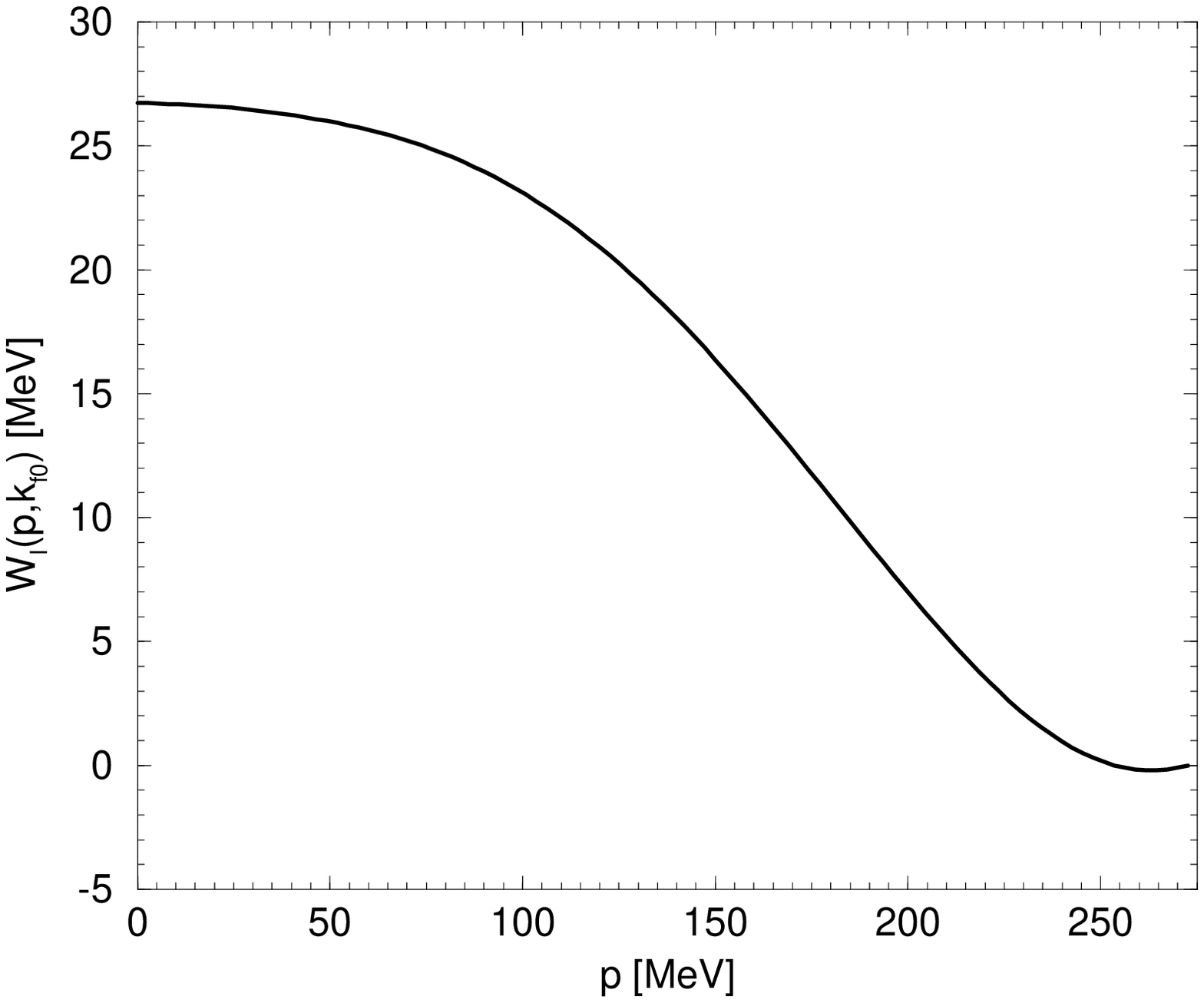}{13.5}
\vspace{-0.8cm}
{\it Fig.\,10: The momentum dependence of the imaginary part of the isovector 
single-particle potential $W_I(p,k_{f0})$ at saturation density
$k_{f0}=272.7$\,MeV.}

\end{document}